\documentclass[final,12pt,3p,times]{elsarticle}

\usepackage{amssymb}
\usepackage{amsmath}
\usepackage{booktabs}
\usepackage{multirow}
\usepackage{rotating}

\usepackage{tikz}
\usetikzlibrary{shapes.geometric, arrows}
\usepackage{comment}
\tikzstyle{startstop} = [rectangle, rounded corners, 
minimum width=3cm, 
minimum height=1cm,
text centered, 
draw=black, 
fill=red!30]

\tikzstyle{io} = [trapezium, 
trapezium stretches=true, 
trapezium left angle=70, 
trapezium right angle=110, 
minimum width=3cm, 
minimum height=1cm, text centered, 
draw=black, fill=blue!30]

\tikzstyle{process} = [rectangle, 
minimum width=3cm, 
minimum height=1cm, 
text centered, 
text width=3cm, 
draw=black, 
fill=orange!30]

\tikzstyle{decision} = [diamond, 
minimum width=3cm, 
minimum height=1cm, 
text centered, 
draw=black, 
fill=green!30]
\tikzstyle{arrow} = [thick,->,>=stealth]

\usepackage{tikz}
\usepackage{pgfplots}
\pgfplotsset{compat=1.17}

\journal{}

\usepackage{xcolor}
\newcommand{\red}[1]{{#1}}
\newcommand{\bluemax}[1]{{#1}}
\usepackage{xcolor}
\usepackage[normalem]{ulem} 

\newcommand{\upd}[1]{\textcolor{red}{#1}} %
\newcommand{\upddel}[1]{\textcolor{red}{\sout{#1}}} 
\renewcommand{\upd}[1]{#1}
\renewcommand{\upddel}[1]{} 
\usepackage{url}
\begin{document}

\begin{frontmatter}



\title{A Realistic Discrete Event Simulation model for Ambulance~Location and Deployment within a regional Emergency~Medical~Service}


\author[label1]{Alberto De Santis} 

\author[label3]{Stefania Iannazzo} 

\author[label2]{Fabio Ingravalle} 

\author[label1]{Stefano Lucidi} 

\author[label2]{Massimo Maurici} 

\author[label1]{Giulia Riccardi} 

\author[label1]{Massimo Roma} 

\author[label2,label3]{Antonio Vinci} 

\affiliation[label1]{organization={Dipartimento di Ingegneria Informatica, Automatica e Gestionale ``A. Ruberti''. \\ SAPIENZA - Università di Roma},
            addressline={via Ariosto 25}, 
            city={Roma},
            postcode={00185}, 
            country={Italy}}

\affiliation[label2]{organization={Dipartimento di Biomedicina e Prevenzione \\ Università di Roma Tor Vergata},
	addressline={Viale Montpellier 1}, 
	city={Roma},
	postcode={00133}, 
	country={Italy}}            
            
\affiliation[label3]{organization={Azienda Regionale Emergenza Sanitaria  - ARES 118},
            	addressline={Via Portuense 240}, 
            	city={Roma},
            	postcode={00149}, 
            	country={Italy}}

\begin{abstract}
The objective of Emergency Medical Services (EMSs) is to promptly respond to calls from citizens for first aid, providing pre-hospital care and, if necessary, to transfer patients to an appropriate Emergency Department (ED) by ambulance. The efficiency of such a system strongly depends on the deployment of ambulance home bases, i.e., locations where ambulances and their crews are strategically positioned, ready to respond to emergency calls. This paper presents a general Discrete Event Simulation (DES) model designed to capture the stochastic behaviour and workflow of regional ambulance emergency systems. 
The proposed model incorporates and integrates information collected from different sources, reproducing very accurately the operation of the ambulance system, thus allowing a more comprehensive and realistic analysis.
To show the applicability and reliability of the proposed general model, a case study provided by the {\em Azienda Regionale Emergenza Sanitaria - ARES 118} (\red{an Italian Regional Emergency Medical Services Authority - ARES~118}) is presented. It concerns a territory within the Lazio region of Italy, including a medium-size city along with sparsely populated areas. The reported results about scenario analyses highlight how the model we propose can be fruitfully used by the managers to improve effectiveness and quickness of the entire regional EMS system.
\end{abstract}


\end{frontmatter}



\section{Introduction}\label{sec1}
Emergency Medical Services (EMSs) play a central role in healthcare delivery, ensuring rapid medical assistance in acute and life-threatening situations. In practice, an EMS provides the prehospital care for a patient according to the following typical steps:
1) to respond to an emergency call and perform the first assessment of the patient conditions;
2) if necessary, to dispatch an ambulance to the emergency scene;
3) to perform patient first treatment on-site;
4) if necessary, to transfer the patient to an Emergency Department (ED) and, after dropping off the patient, return to the base or head to another emergency scene.
\par
Therefore, the effectiveness of EMS largely depend upon the operational efficiency of ambulance services, as a timely arrival to patient location is essential in order to initiate early care and stabilize clinical conditions before hospital treatment can begin. Hence, ambulances represent one of the most important resource within EMS and the main Key Performance Indicator (KPI) is the Response Time (RT), i.e., the time interval between an emergency call being received and an ambulance arriving to the emergency scene. RT is critical for the effectiveness of prehospital care, e.g., medical interventions provided to patients at the scene and during transport to an ED. In time critical conditions (e.g., cardiac arrest, major trauma or stroke) even a small reduction of the RT can substantially increase survival probability and improve clinical outcomes \red{(see, e.g. \cite{aringhieri.2017})}.
\par 
A key factor strongly affecting RT is the location of the ambulance home bases, the deployment of ambulances among the bases and the dispatching of ambulance to the emergency scene. This motivates the great attention devoted
to ambulance location and deployment, namely to the following two issues:
\begin{itemize}
	\item ambulance home bases location (\textit{location problem})
	\item ambulance deployment among the bases (\textit{deployment problem}).
\end{itemize}
Note that if the vehicles are identical, ambulance deployment reduces to determine the number of ambulances assigned at each base.
\par
Bases are the locations where ambulances and their crews are strategically positioned within a territory ready to respond to emergency calls. Even if the locations of these bases have a significant impact on the quality and effectiveness of prehospital (and potentially also hospital) care, in practice, ambulance base locations are often determined only according to some generic guidelines or administrative convenience rather than by systematic data driven analysis. That is, usually, bases tend to be located in some sites without a rigorous evaluation about some criterion related to current demand patterns. The same holds for ambulance deployment among the bases.
\par
It is worth noticing that the size of a deployment problem for realistic instance is very large. In fact, if we consider $n$ identical ambulances to be deployed in $m$ bases, each of them contain any number of ambulances, the resulting deployments are $\binom{n+m-1}{n}$. Therefore, if ambulance emergency system performance evaluation is time consuming (as when simulation is used), any strategy based on the comparison of all the deployments can not be employed.
\par
In this paper we introduce a general purpose and reusable Discrete Event Simulation (DES) model capable of representing the operational dynamics of ambulance emergency system, aiming at improving its overall effectiveness by means of an enhanced base location and deployment of the ambulance fleet in a region. This general model enables to simultaneously consider both location and deployment problem and it can be used for both scenario analysis to compare the performance of alternative configurations and within the Simulation Based Optimization (SBO) approach to identify configurations that optimize some KPIs. 
%
\par
It is important to highlight that the proposed modelling framework may need to be adapted to reflect the EMS system under study's specific characteristics, including its dispatching protocols, fleet composition and hospital network configuration. Nevertheless, the model we propose provide a general and flexible framework that can easily be customized for different operational contexts. 
\par
The distinguishing features of the proposed model lie in a more accurate and realistic representation of ambulance system operations compared with existing models in the literature. Recent survey papers on ambulance location and deployment problems
(see the very recent paper \cite{karpova.2026}) highlight that many existing models rely on simplifying assumptions that limit their ability to represent real EMS operations.
Therefore, in the last years, research in DES applied to ambulance location problem mainly focused on building models that reproduce the entire ambulance service process as realistically as possible.
Our work addresses the limitations existing in many models through a realistic DES framework that explicitly models the full operational path of an ambulance mission. In particular, this is achieved by weakening some simplifying assumptions commonly adopted in the literature.
\par
For instance, we incorporate operational delays that are frequently neglected in the literature \cite{frichi.2022, sudtachata.2016}, such as telephone triage and ambulance preparation times, which are modeled using empirical distributions estimated from historical data. As stated in \cite{karpova.2026} in their review, 66\% of studies considered either set triage time to zero or not mention it, and 78\% disregarded preparation time. These omissions can lead to a significant overestimation of the effectiveness of relocation strategies, as demonstrated by the tests carried out.
\par
Moreover, many models do not explicitly consider times due to ambulance offload delay \red{and ambulance ramping\footnote{\red{The term {\em ``ambulance ramping''}
(see, e.g., \cite{kingswell.2017}) means that patients arriving at EDs via ambulance cannot be transferred into the ED’s care in a timely manner and the ambulance is ``ramped''. It does not mean the patient is still in the back of the ambulance, just that the transfer of care from ambulance crew to the ED clinicians has not yet happened.}}}, i.e. the possibility that an ambulance arrives at the ED but no bed or staff are available to take over the patient \cite{asplin.2003,mclay.2010}. Hence, \red{healthcare professionals} must stay with the patient (sometimes for hours), keeping the ambulance unavailable for new emergencies. Since this phenomenon occurs very frequently, it is important to consider this time separately, in order to monitor how long ambulances remain unavailable while waiting at the ED. In \cite{karpova.2026} is stated that in 47\% of studies considered, handling time 
at the hospital is omitted. Handling time is not part of the RT, but it is of great importance because it causes ambulances to be occupied for longer with one patient, avoiding their availability for others emergencies, see, e.g., \cite{aringhieri.2017,li.2019,eckstein.2004}.
\par
Another model enhancement concerns the use of more sophisticated criteria for choosing the ED where the patient is delivered. In fact, in our model, the choice is not only simply based on proximity criteria (typically the nearest ED) \cite{karpova.2026}, but on the diagnosed pathology, as actually performed in practice. This is possible by referring to different levels of hospital-based emergency care capacity within an integrated referral network. This approach is consistent with the WHO hub-and-spoke and tiered emergency care models nowadays widely adopted in National Healthcare Systems (NHSs) \red{\cite{WHO.2019}}, including the Italian one.
\par
Furthermore, we adopt a flexible demand generation mechanism with spatial hierarchy, capturing both temporal variability and fine-grained spatial heterogeneity of emergency calls. Finally, travel times are computed on the actual road network and calibrated using historical observations, thus avoiding the common simplifying assumptions of Euclidean distances or constant speeds \cite{maxwell.2009,sudtachata.2016}.
In \cite{alanis.2013} the authors study how the type of distance (estimated with the road network or Euclidean) affects the probability of the response within a certain time, indicating that this probability increases with the Euclidean distance.
\par
In addition, the model we propose also takes into account possible changes of the severity code assigned to patients on site, i.e., once the ambulance reached the scene
\cite{kim.2023}. This is not an infrequent occurrence, since the first triage (when the initial severity code is decided) is usually performed during the emergency call, without a direct patient examination. 
\par
As concerns the performance measures, we refer to the already mentioned RT which represent the most significant timeliness metric. In particular, following recent literature (see, e.g., \cite{aboueljinane.2023}), we focus on the so called \textit{coverage}, that is the percentage of calls for which the RT does not exceed a prefixed target time. Several other KPIs can be defined also related to survival rate and financial aspects (see \cite{aboueljinane.2013A,zaffar.2016}).
\par
Overall, the proposed framework provides a high-fidelity simulation model capable of capturing several operational features that are typically neglected in EMS location models.
\par
To show the reliability of the proposed general model, we cosidered a real case provided by the {\em Azienda Regionale Emergenza Sanitaria - ARES 118}
concerning a territory within the Lazio region of Italy, including the medium-size city of Rieti 
along with its province which also consists of sparsely populated areas.
The rationale behind this choice is to consider a wide area with different population density so that the \red{emergency} calls are expected to be variously distributed. The availability of a large amount of data allowed to build a very accurate model well reproducing the ambulance emergency system in that region and to perform valid and realistic scenario analyses.
\par
The model has been implemented by using {\sf SimPy} \cite{simpy}, a process-based DES framework based on standard Python\footnote{https://simpy.readthedocs.io/en/latest/}. Validation results showed a close agreement between simulated and real system, so that the model can be successfully adopted as an accurate representation of the EMS under study.
The scenario analysis performed on the real case showed that decisions concerning ambulance deployment on a territory should explicitly account for demand concentration, temporal availability of resources, and the interaction between spatial positioning and stochastic system dynamics. 
\par
Overall, the study we performed highlighted the importance of adopting a realistic DES model, when dealing with ambulance location and deployment problem. The model we proposed revealed very accurate, well reproducing the complex system dynamics of an EMS. It can be successfully adopted by EMS planner as reliable decision support tool also to ensure efficinency and equity in the services provided.
\par
The paper is organized as follows: Section~\ref{sec2} provides a review of the relevant literature on ambulance deployment in EMSs. Section~\ref{sec3} presents the proposed DES framework, detailing its structure, assumptions, and modeling choices. Section~\ref{sec4} introduces the case study and describes the data, as well as the model calibration and validation procedures. It also reports the results of the scenario analysis. Finally, Section~\ref{sec5} reports some concluding remarks and directions for future research.

\section{Literature review}\label{sec2} 
The management of ambulance fleets within EMS has been subject of extensive research across the domains of Operations Research, Simulation, and Artificial Intelligence. The great interest in ambulance location problem within an EMS system is evidenced by the very large number of papers published on this topic over the past five decades. The reviews \cite{revelle.1977,botcorne.2003,goldberg.2004,li.2019,basar.2011,pinto.2015,aringhieri.2017,belanger.2019,Frichi.2024,neira-rodado.2024, karpova.2026} report a great number of references and we refer the reader to them for a complete literature review.
%
In the sequel of this section we mention some papers representative for each approach proposed in literature.
\par
Models dealing with ambulance location problem can be classified in \textit{static} or \textit{dynamic} and \textit{deterministic} or \textit{stochastic} and contributions have evolved from static and deterministic models to dynamic, data-driven, and hybrid approaches capable of capturing the inherent stochasticity of EMS.
\par
One of the earliest methodological approaches to ambulance deployment relies on {\em static deterministic models} that use optimization methods for solving static ambulance siting and allocation problems. These models typically aim at identifying the best configuration of bases and vehicles that maximizes the covered demand or minimizes travel time (or number of vehicles). Two classical formulations are the Location Set Covering Problem, proposed in \cite{toregas.1971}
and the Maximal Covering Location Problem, proposed in \cite{church.1974}.
Unfortunately, these models do not take into account possible unavailability of a vehicle of responding to a call since already busy in a previous task. Therefore 
subsequent extensions introduced \textit{stochastic models} considering the probability that an ambulance is available on emergency call arrival. One of the first paper incorporating this issue is \cite{daskin.1983} where a Maximum Expected Covering Location Problem (MEXCLP) is proposed. 
\par
Static models, though useful for strategic planning, fail to capture real-time system dynamics. The notion of dynamic redeployment, i.e., the repositioning of idle ambulances to back up busy ones represents a major step forward in EMS research \cite{maxwell.2009}.  
Two broad categories can be distinguished: multi-period models, which prescribe vehicle locations at discrete time intervals, and fully dynamic models, which continuously adapt to the system state.  
Later works have increasingly emphasized real-time and dynamic decision making. For instance, \cite{mason.2013} proposed the Real-time Multi-view Generalized-cover Repositioning Model, balancing service quality and relocation costs, while \cite{vanbarneveld.2015} introduced the Dynamic Maximum Expected Covering Location Problem (DMEXCLP) to minimize the fraction of late arrivals.  
Other authors have expanded the scope of optimization to multi-scenario or stochastic frameworks, such as \cite{nickel.2016}, who minimized costs under coverage constraints, and \cite{boujemaa.2018}, who designed a two-stage stochastic program simultaneously determining the location, composition, and assignment of the EMS fleet.  
A comprehensive overview is provided by \cite{grekousis.2019}, who classified EMS location models into three families according to their objective functions: (i) covering problems, which trade off fleet size and covered population; (ii) center problems, minimizing the maximum distance between demand and the nearest base; (iii) median problems, minimizing the average travel distance or time.  
\par
Dynamic extensions of classical covering models have been proposed by \cite{gendreau.2001}, \cite{moeini.2014}, \cite{basar.2011}, and \cite{schmid.2012}, who introduced penalties for excessive relocations, variable travel times, and stochastic demand. More sophisticated models rely on Approximate Dynamic Programming \cite{maxwell.2009, schmid.2012} to manage large state spaces efficiently, while \cite{naoum.2013} and \cite{rajagopalan.2008} addressed uncertainty through stochastic and queueing-based formulations. Despite significant progress, as noted by \cite{belanger.2016}, comparative evaluations of redeployment strategies remain scarce, and the search for robust, computationally efficient dynamic policies continues to be an open research frontier.
\par
Recognizing the limitations of purely analytical models, \textit{dynamic models based on DES} have been proposed and widely currently used to study ambulance allocation problem. Indeed, they provide an effective tool for dealing with the problem in a more realistic way. They enable the reproduction of the sequence of stochastic events that characterize ambulance emergency systems (e.g., call arrivals, dispatching, travel to the scene, patient transport and hospital admission) and provide an experimental environment to perform significant scenario analyses. For a review on simulation models for EMS we refer to the paper \cite{aboueljinane.2013A}. 
\par
A noteworthy paper dealing with ambulance deployment based on DES is \cite{aboueljinane.2014} (see also \cite{aboueljinane.2013B,aboueljinane.2023}). In this paper, an enhanced simulation model is adopted to replicate the French national EMS under general assumptions. The reported results
demonstrates how simulation can provide an effective tool for assessing deployment and dispatching policies. Similarly, \cite{henderson.2005} employed simulation to assess whether routing and dispatching decisions conform to predefined operational rules, while \cite{ingolfsson.2003} analyzed how stochastic variations in demand and travel time affect key performance indicators such as average response time and server utilization.  
Earlier studies, including \cite{lubicz.1987} and \cite{repede.1994}, established the importance of validating simulation outcomes against real data, laying the methodological foundations for modern DES applications in EMS planning.

Recent research has proposed to merge simulation and optimization in the SBO unified framework (see, e.g., \cite{fu.2015}). It exploits optimization to identify candidate solutions and simulation to evaluate KPIs of interest under realistic stochastic conditions, in an iterative loop. The combined use of the two approaches in solving the ambulance location problem also arises from the necessity to assess results obtained by optimization method. \cite{aboueljinane.2023} applied such an approach to the French EMS, formulating a multi-period linear program solved with CPLEX optimizer and then validated through simulation to account for different demand and priority scenarios. In \cite{aringhieri.2016} the authors reported a study based on statistical modelling, simulation and mathematical programming on the ambulance location and management in Milan (Italy) EMS, showing that it is possible to improve the quality of the delivered service by using these analytic techniques.  
Likewise, \cite{zhen.2014} combined an integer linear program and a genetic algorithm to optimize relocation costs, with simulation used for performance testing. \cite{nogueira.2016} extended this paradigm by incorporating operational constraints such as depot activation and capacity limits, explicitly modeling the trade-off between service times and operational costs.  
These integrated models demonstrate that coupling simulation and optimization can provide decision makers with solutions that provide optimal and viable solutions, offering a valuable balance between precision and realism.
\par
Furthermore the literature identifies numerous issues that influence ambulance bases positioning and criteria for selecting a hospital where the patient is delivered. \cite{marsh.1994} first linked population density to coverage optimization; \cite{maleki.2014} and \cite{vanbarneveld.2016} incorporated vehicle unavailability; \cite{schmid.2012} accounted for congestion and time-dependent travel times and \cite{aboueljinane.2013B} emphasized multi-vehicle and multi-priority systems.  
The considered KPI vary widely, from total service costs \cite{ball.1993} to time between call and dispatch \cite{ingolfsson.2003}, and have progressively shaped the set of KPIs typically used to evaluate EMS logistics (see also \cite[Section 2.3]{aboueljinane.2013A} and \cite{zaffar.2016}). 
\par
More recent papers are devoted to the use of metamodels (or surrogate models), In fact, as well known, to capture the full complexity of an EMS system by means of DES typically requires a very high computational burden to generate credible results. Metamodels represents a valid alternative to this inconvenient. They construct a simple approximation of the input/output function implicitly defined by the underlying simulation model. Then, the analytic expression of this approximation is used in an optimization process which provides solutions in drastically reduced computer time. A recent review on simulation based metamodelling in EMS can be found in \cite{sahlaoui.2023}.
%
\par
Finally, we mention how the rapid development of Artificial Intelligence has recently opened new perspectives for EMS decision support systems. Artificial Intelligence techniques can enhance both situational awareness and proactive decision making by enabling real-time monitoring, anomaly detection, and demand forecasting.  
As highlighted by \cite{aringhieri.2017}, EMS operations generate vast amounts of historical and real-time data (e.g., travel times, hospital occupancy, etc.) that can be exploited to predict future demand \cite{chen.2016, huang.2019, jin.2021, lee.2021, martin.2021}. Predictive analytics thus facilitates the dynamic pre-deployment of ambulances, reducing response times and improving patient outcomes \cite{grekousis.2019}. We refer to \cite{Tluli.2024} for a comprehensive survey covering ambulance allocation, routing, demand prediction, and deployment methods with emphasis on Machine Learning-based methods in EMS planning.
\par
Concluding this literature review section, it is worth mentioning that, in the last two decades, some EMS simulation software packages have also been developed. The most recent is the JEMSS package \cite{ridler.2022}, a free and open-source EMS simulation and optimization package developed in the {\sf Julia} programming language. Interested readers are referred to \cite{ridler.2022} for a comprehensive overview of available EMS software packages.

\section{A general purpose DES model for ambulance location problem}\label{sec3}
The literature review highlights that simulation models are widely used to evaluate ambulance deployment and relocation strategies before their implementation in real-world EMS systems. By incorporating operational delays, actual road networks, and non-homogeneous call arrival rates, these models provide a more realistic representation of system operations and reduce the risk of overestimating performance and response capacity.
This motivates the introduction of improved, comprehensive and more realistic simulation models, able to capture the stochasticity and the dynamics which rule the complex ambulance emergency system. This level of realism increases the credibility of the evaluation tools and facilitates their adoption by EMS managers.
Based on this observation, we developed a general purpose and \red{flexible (reusable)} DES model that incorporates and integrates  the collected information, aiming at reproducing very accurately the actual operation of the ambulance emergency system.
\par
In this section, starting from a  problem description, we detail the general DES model we propose, highlighting the main distinguishing features. In the subsequent Section~\ref{sec4} we apply this model to a real case study, showing its usability and reliability.

\subsection{Problem description}\label{sec:pbdescr}
We briefly describe all the typical processes that characterize an ambulance emergency system, involving several and different types of human and material resources. Even if EMS may vary from a country to another, several standard components and procedures can be evidenced (see, e.g., \cite{pinto.2015}). In the sequel we report the general scheme we adopt, highlighting some distinguishing peculiarities which will be then included in our enhanced DES model.
\par
Emergency call arrives at the Operations Center where an operator answering the call collects information on the request and decides the appropriate response. In particular, after discharging possible inappropriate calls, the operator assigns a severity level according to the perceived seriousness and urgency (telephone triage).
Then, if the operator decides that an ambulance must be dispatched, a selection rule is adopted to assign an ambulance (and its crew) to the calling patient. In particular, the operator evaluates the geographical position of the patient with respect to the nearest available vehicle. Once the assignment is finalized, the ambulance undergoes a travel from its base (or current position) to the scene of the emergency. Ambulances always take the fastest path available to an emergency scene and if the estimated travel time to reach the scene exceeds a predefined threshold, the ambulance may not be dispatched and the request is instead queued until a closer ambulance becomes available to respond.
\par
The intervention of emergency personnel at the scene can be carried out in \red{three} ways: either it \red{terminates during the travel to the scene since the mission is cancelled} or it concludes on site after a treatment (for minor cases when no transport is required) or patient requires transport to an ED (possibly after stabilizing his/her conditions). In the latter case, an ED must be selected not only according to the shortest path criterion, but also \red{taking} into account \red{the patient's} (\red{presumptive}) diagnosed pathology. In fact, according to the
\red{prevalent Hub-and-Spoke framework \cite{bekelis.2019,ginzberg.2023},}
 it is possible to distinguish 
\begin{itemize}
	\item hospitals with essential emergency care capacity for undifferentiated acute conditions managing low to moderate complexity emergencies that refer patients requiring advanced or highly specialized care (\textit{spoke}); 
	\item hospitals with comprehensive and highly specialized emergency care capacity for high-acuity, time-critical, and complex conditions, ensuring definitive care for life-threatening conditions (\textit{hub}).
\end{itemize}
Therefore, based on patient severity conditions and specific pathology, the nearest suited ED is selected and the ambulance start the \red{travel} from the scene to the chosen ED. Upon arrival, a handover phase takes place, during which the patient should be transferred to the ED.
Unfortunately, a common and critical phenomenon at this stage is \textit{Ambulance Offload Delay (AOD)}. This occurs when, due to emergency department (ED) overcrowding, no treatment spaces, stretchers, or ED personnel are immediately available to receive patients arriving by ambulance. Consequently, patients cannot be transferred promptly through the ED, resulting in delays in definitive care (for a review on AOD, see, e.g., \cite{li.2019}). When AOD occurs, ambulance crew wait with the patient, and continue to provide patient care until an ED bed becomes available. 
Therefore the delayed ambulance and crew are unable to return to service (ambulance \red{ramping}), and this delay can be significant, having an adverse impact on ambulance emergency system.
\red{Studies adopt different threshold values} as benchmark for delayed offload, i.e. time interval between ambulance arrival at the ED and when patient care is formally transferred from the ambulance crew to ED staff \red{(see, e.g. \cite{eckstein.2004,cone.2012,cooney.2013}).}
\par
The ambulance mission is completed when the patient handover is accomplished. After mission completion, the ambulance may require sanitization before it can return to service; this process usually takes place at its own base and temporarily removes the vehicle from service. Conversely, if no sanitization is needed, the ambulance can either return to its base or be reassigned directly to another emergency.       
\par
\begin{figure}[htbp]
\centering
\begin{tikzpicture}[node distance=1.75cm]

\node (start) [startstop] {Call arrival};
\node (pro1) [process, below of = start] {Telephone Triage};
\node (pro2) [process, below of = pro1] {Ambulance dispatch};
\node (pro3) [process, below of = pro2] {Travel to the scene};
\node (dec1) [decision, below of = pro3, yshift = -0.5cm] {ED ?};
\node (pro4) [process, right of = dec1, xshift = 2cm] {Treatment on site};
\node (pro5) [process, below of = dec1, yshift = -0.5cm] {Load of the patient on the ambulance};
\node (pro6) [process, below of = pro5] {Travel to the ED};
\node (dec2) [decision, below of = pro6, yshift = -0.5cm] {AOD ?};
\node (pro7) [process, right of = dec2, xshift = 2cm] {Patient discharge from the ambulance};
\node (pro8) [process, below of = dec2, yshift = -0.5cm] {Ambulance block};
\node (dec3) [decision, right of = pro5, xshift = 7cm] {Sanitization?};
\node (pro9) [process, below of = dec3, yshift = -1.9cm] {Return to base};
\node (pro10) [process, below of = pro9, yshift = -0.9cm] {Sanitization};
\node (dec4) [decision, right of = pro3, xshift = 7cm ] {Call nearby?};
\node (pro11) [process, right of = pro1, xshift = 7cm ] {Return to base};
\node (stop) [startstop, right of = start, xshift = 7cm ] {Ambulance available};

\draw [arrow] (start) -- (pro1);
\draw [arrow] (pro1) -- (pro2);
\draw [arrow] (pro2) -- (pro3);
\draw [arrow] (pro3) -- (dec1);
\draw [arrow] (dec1) -- node[anchor = south]{no} (pro4);
\draw [arrow] (dec1) -- node[anchor = west]{yes} (pro5);
\draw [arrow] (pro5) -- (pro6);
\draw [arrow] (pro6) -- (dec2);
\draw [arrow] (dec2) -- node[anchor = south]{no} (pro7);
\draw [arrow] (dec2) -- node[anchor = west]{yes} (pro8);
\draw [arrow] (pro8) -| (pro7);
\draw [arrow] (pro4) |- (dec3);
\draw [arrow] (pro7) |- (dec3);
\draw [arrow] (dec3) -- node[anchor = west]{no} (dec4);
\draw [arrow] (dec3) -- node[anchor = west]{yes} (pro9);
\draw [arrow] (pro9) -- (pro10);
\draw [arrow] (dec4) -- node[anchor = west]{no} (pro11);
\draw [arrow] (dec4) -- node[anchor = south]{yes} (pro3);
\draw [arrow] (pro11) -- (stop);
\draw[arrow] (pro10.east) -- ++(0.8,0) |- (stop.east);
\end{tikzpicture}
\caption{Flowchart of the ambulance emergency system processes}
\label{fig:model}
\end{figure}
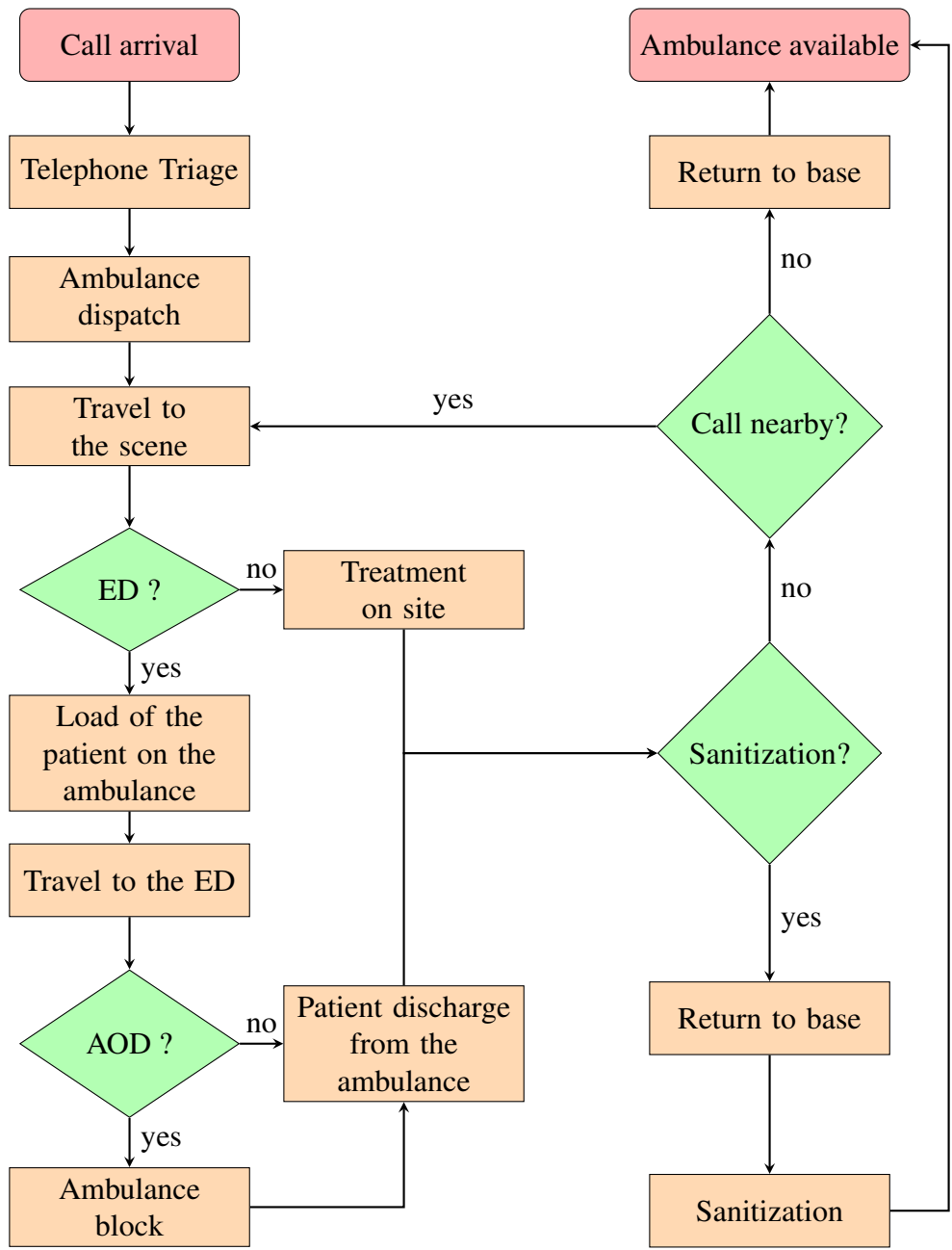

It is also possible for a mission to be cancelled at any stage, for example, if another ambulance reaches the patient first. In such cases, ambulance mission terminates prematurely. A flowchart summarizing the processes characterizing ambulance emergency system is reported in Figure~\ref{fig:model}.
\par
Ambulances can be classified according to the level of medical care provided and their service schedule. Typically, two types of ambulances can be used \red{\cite{isenberg.2005}}: 
\begin{itemize}
    \item \textit{Basic Life Support} (BLS) ambulances that provide basic pre-hospital care and are usually staffed by trained emergency personnel. They perform non-invasive interventions such as monitoring vital signs, oxygen administration, cardiopulmonary resuscitation, and the use of automated external defibrillators. These units are generally used for non-critical emergencies or patient transport where advanced interventions are not required.
    \item \textit{Advanced Life Support} (ALS) ambulances that are designed to manage high-acuity medical emergencies typically staffed by emergency physicians, or nursing staff trained in advanced prehospital care. They are equipped to perform advanced procedures, including intravenous drug administration, airway management (e.g., intubation), and electrocardiogram interpretation.
\end{itemize}
Depending on the severity of the emergency, dispatch may involve a BLS ambulance alone or both BLS and ALS ambulances. The specific dispatch policy may vary across EMS systems and local protocols.
Furthermore, ambulances may operate under different shift schedules, typically H12, corresponding to daytime coverage (12 hours), and H24, corresponding to continuous day-and-night coverage (24 hours). Consequently, the effective availability of ambulances in the EMS system depends not only on the number of vehicles but also on the shift schedule assigned to each ambulance.


%
%
%
\subsection{Assumptions}
We now itemize the assumptions adopted in our model. Most of them are common to the models already proposed in literature:
\begin{itemize}
    \item  restocking, crew shifts and maintenance are not considered;
	\item travel times are assumed to be deterministic, but a detailed time-varying travel is used, depending upon time slot of the day and urgency level;
		\item the \textit{ambulance diversion} is not allowed, i.e., the strategy of temporarily redirect patients incoming by ambulance to another nearby ED when the receiving ED is overloaded (see, e.g., \cite{AD-TechRep.04-2025} and therein references);  
	\item \textit{pre-emption priority} in dispatching a rescue team is not permitted, i.e., the possibility of reassigning an ambulance serving a low priority call to a higher priority call;
	\item \textit{on-idle redispatch} is not allowed, i.e., an ambulance assigned to an emergency call can not be changed since another ambulance closer to the location of the call has become available;
	\item \textit{on-road redispatch} is not allowed, i.e., an ambulance can not be dispatched to a new call during the returning travel to its own base.
\end{itemize}
Note that most of these assumptions are not actual limitations, since they closely reflect the regulation adopted by many regional EMS. For instance, 
some states or regions have restricted or eliminated ambulance diversion (mentioned at the third item) because possible delay of life-saving treatment. Anyhow, when detailed operational data are available, ambulance diversion policies can be easily incorporated into the simulation model we propose with limited modifications to the modelling framework.
\par
\red{Moreover, as regards the first assumption, it concerns the support activities that take place during phases when the vehicle is not in service, so that it does not actually affect the ambulance system process.} The last assumption of the list implies that the direct graph underlying the considered road network consists of nodes corresponding to: ambulance basis, emergency scenes, ED locations. Even if this precludes modelling more complex behaviours, it characterizes almost all the DES models for EMS proposed in literature. This is due to the difficulty in determining ambulance locations on the return trip for redispatching. As far as the authors are aware, the only papers which propose a DES model which includes this redispatching feature are \cite{ridler.2022} and \cite{janosikova.2019}. In the first paper, this is obtained by introducing many additional intermediate points to the road network and then considering a ``reduced network'' with respect to the entire network so that the shortest path between all pairs of nodes can be more efficiently calculated. Anyhow, even if the considered network is shrunk, very large instances can hardly be handled. In the second paper, the authors do not explicitly describe how this feature is implemented in their model. 
\par
More generally, it is important to highlight that the proposed modelling framework may need to be adapted to reflect the specific characteristics of the EMS system under study, including its dispatching protocols, fleet composition and hospital network configuration. \red{Finally, it is important to highlight that any changes in ED network configuration, e.g. a temporary unavailability  of some EDs or  a reduced capacity, can be easily included in the model, provided that sufficient information is given.}

\subsection{Input data}\label{sec:data}
An accurate data collection is at the basis of any DES models. Data required to build a valid model for the full  workflow of an ambulance emergency system
are the following:
\begin{itemize}
	\item area of interest and characteristics (extension, population density, etc.);
	\item geographical location of the current ambulance bases and EDs in the area, along with the current deployment of ambulances among bases;
	\item inter-zone travel times for each route and time period (day, night, peak hours, etc.);
	\item call arrivals with spatial and temporal indications, assigned severity code, \red{presumptive} pathology;
	\item dispatching rules;
	\item times to perform tasks (receiving a call, rescue team preparation, treatment on the scene, drop off at ED with possible AOD);
	
	\item the level of care and specialist services available at each ED, along with predefined territorial referral rules; 
	
	\item sites where additional ambulance bases could be located and number of possible additional ambulances to deploy.
\end{itemize}
The last point concerns the scenario analyses when it is required to investigate about additional bases and/or ambulances.

\subsection{Model description}\label{sec:modeldescr}
The overall structure of the model reflects the flowchart of the ambulance emergency system processes reported in Figure~\ref{fig:model}. The model entities are represented by \textit{the emergency calls} (temporary entities) and the \textit{ambulances} (permanent entities). At the beginning of the simulation all the ambulances are generated and located at their own base. 
Emergency calls are then generated according to the call arrivals scheme.
At each entity (ambulance and call) attributes reporting all the needed information are assigned. For the emergency calls that require an ambulance sent to the emergency scene, a dispatching rule is adopted for choosing the most suited vehicle. Then for the selected ambulance the following task are scheduled: preparation of the rescue team, travel to the scene, treatment on site,  delivery to an ED, patient drop off including possible AOD, return to the base or direct dispatch to another emergency scene (if sanitization at base is not necessary). Of course an ambulance mission may terminate on the scene if transport to an ED is not necessary. 
\par
In the sequel we detail all the aspects which characterize the model.

\subsubsection{Call Generation}\label{sec:gencall}
To reproduce the demand for EMS, each simulated call is defined by two key attributes: the time of occurrence and the spatial location corresponding to the patient position.
The spatial component is modelled using a hierarchical structure that subdivides the area under study into two complementary spatial layers:
\begin{itemize}
	\item\textit{generation zones}: relatively large territorial units, each containing a sufficiently high volume of historical calls to permit an accurate estimation of a zone-specific call interarrival time distribution;
	\item \textit{call squares}: the entire area under study is further discretized into smaller, uniformly sized subareas (for instance, squares of approximately 5-10 km$^2$) whose extension depends on the area considered. 
\end{itemize}
The call generation process follows a two-stage mechanism.
First,  a call arrival time is drawn from the interarrival distribution of a certain call generation zone.
Second, within the chosen generation zone, a specific call square is randomly selected, again according to the historical spatial density of calls.
Within the selected square, the exact patient location is determined by sampling a random coordinate, which can correspond either to the centroid of the square or to a random point extracted from the empirical dataset of past emergencies.
The temporal component of call generation is modeled through an analysis of historical demand patterns. The call arrival rate is studied as a function of several temporal dimensions, including hour of the day, day of the week, month, and season. Only those temporal variables that show a statistically significant influence on call frequency are retained in the model. For instance, if no substantial variation is observed across months, the monthly factor is omitted, simplifying the model without compromising accuracy.
\par
The call square size should be small enough to ensure an adequate spatial resolution of emergency scene localization, yet not so small as to excessively increase model complexity.
Since each square typically covers a limited area and may contain only a small number of historical calls, it is often impractical to estimate a specific interarrival time distribution for each of them.
For this reason, the hierarchical two-stage mechanism is adopted: generation zones aggregate sufficient call volumes to allow the estimation of reliable interarrival distributions, while call squares capture the finer spatial heterogeneity of demand within each zone.
This structure balances statistical robustness with spatial accuracy, ensuring that both temporal and geographical patterns of EMS demand are realistically reproduced in the simulation.

\subsubsection{Travel times}
\label{subsec:travel-times}
At this stage, each simulated call is associated with a set of spatial coordinates: those of the emergency scene location (determined as described in the previous section), the nearest appropriated ED, and the ambulance bases (or candidate base locations).
To estimate travel times between these points, a routing services such as the {\sf OpenRouteService} API\footnote{It is opensource and freely available at 
{\tt openrouteservice.org}}
or similar geospatial tools capable of computing realistic road-network distances and travel durations can be adopted.
This approach is substantially more accurate than using simple Euclidean distances, as it accounts for the actual street network configuration, including road hierarchies, intersections, and travel constraints.
The travel times returned by routing services should be interpreted as nominal values, as they do not explicitly account for operational factors such as traffic congestion or emergency driving conditions (e.g., sirens and priority lanes). 
\par
To address this limitation, it is convenient to introduce a multiplicative calibration factor which can be estimated from historical data, distinguishing among time slot \(t\) (e.g., day/night, peak/off-peak) and urgency class \(u\) (e.g., urgent vs. non-urgent missions).
\par
For each combination \((t,u)\),
let \(T_{\text{RS},r}^{(t,u)}\) denote the travel time for the route
\(r\) estimated by the routing service (RS), and let \(T_{\text{OBS},r}^{(t,u)}(s)\), 
$s \in \mathcal{S}_r(t,u)$ be the corresponding travel times observed (OBS) in historical mission data, for the same route $r$, where \(\mathcal{S}_r(t,u)\) denotes the set of observations for the route $r$, associated with time slot \(t\) and urgency level \(u\) (after applying consistency filters to data to remove extreme or implausible discrepancies between routing-based and observed travel times). The scaling factor \(\alpha_r(t,u)\) for the travel time of the route $r$, associated with time slot \(t\) and urgency level \(u\), is determined as
\[
\alpha_r(t,u) \;=\; \arg\min_{\alpha > 0} 
\sum_{s \in \mathcal{S}_r(t,u)}
\left| \alpha \, T_{\text{RS},r}^{(t,u)} - T_{\text{OBS},r}^{(t,u)}(s) \right|,
\]
and the calibrated travel time used by the simulation model (SIM) is
$
T_{\text{SIM,r}}^{(t,u)} = \alpha_r(t,u) \cdot T_{\text{RS},r}^{(t,u)}.
$
\par
The calibration procedure is applied separately to different operational legs of the mission (e.g., base-to-scene, scene-to-scene, scene-to-ED), allowing the model to capture systematic differences in travel dynamics across mission phases. When historical data for a route $r$ are insufficient for a given combination \((t,u)\), a conservative default value \(\alpha_r(t,u) = 1\) is adopted, corresponding to the use of nominal routing times without correction.
When sufficient historical data are available, the model can be further refined by introducing zone-specific correction indices, capturing spatial heterogeneity in traffic patterns (e.g., urban vs. rural areas). 
This calibration process enables the simulation to reproduce realistic travel times under varying temporal and urgency conditions, while maintaining computational efficiency and consistency with real-world observations. After computing the estimated travel time $t$ (in minutes), when the exact location of the emergency is not precisely known (e.g., when square grid cells are used), additional variability is introduced. Specifically, a random value is drawn from a triangular distribution with parameters $(t-\delta,t,t+\delta)$, where $t$ is the estimated travel time and $\delta$ represents the maximum deviation from the nominal value. Negative values are truncated at zero. This approach allows us to capture the uncertainty associated with approximated emergency locations.

\subsubsection{Service times}\label{sec:servicetimes}
The following service times are modeled in the simulation model we propose: 
\bluemax{
\begin{itemize}
\item telephone triage 
\item ambulance assignment 
\item ambulance preparation
\item treatment on site 
\item load of the patient on the ambulance
\item ambulance \red{ramping} at the ED (if any)
\item patient discharge from the ambulance
\item sanitization (if any).
\end{itemize}}
Each of these time components is represented by means of a probability distribution estimated from historical data. The estimation process distinguishes between emergency cases (high severity codes)
and non-emergency cases (low severity codes)
to capture the different operational dynamics and durations associated with each urgency level. The urgency levels considered before the ambulance arrives on scene correspond to the priority codes assigned during the telephone triage phase. Once the ambulance reaches the scene and the patient is clinically assessed, the urgency level may be revised; from that point onward, the urgency codes assigned on scene are used in the model. In the case of AOD at an ED, the probability distribution is further distinguished by ED, as patient offloading and turnaround times may vary considerably depending on the facility’s capacity, organization and congestion level.
\par
A theoretical probability distribution for each service time should be obtained by fitting the data, estimating the parameters and using goodness-of-fit tests (Kolmogorov-Smirnov, Chi-squared, Anderson-Darling tests).
When the tests indicate a lack of fit
(rejection of the null hypothesis), an empirical distribution should be adopted.

\subsubsection{ED selection}
EMS typically routes patients to ED based on clinical suitability and proximity. The destination is chosen by matching the patient’s condition with the level of care and specialist services available, following predefined territorial referral rules that link each area to specific facilities.
As already mentioned in Section~\ref{sec:pbdescr}, 
according to the modern WHO tiered emergency care models, emergency networks usually differentiate between general hospitals, advanced EDs and specialized hubs for time-sensitive conditions such as stroke, trauma or cardiac emergencies
\cite{agenas.2018}. Operational factors such as distance, travel time and type of responding ambulance, are also considered, together with real-time information on crowding to guarantee timely and adequate access to definitive care.
In the simulation model, this mechanism can be abstracted by assigning at each emergency a territorial reference with a given probability and routing the patient to the nearest ED within that reference.

\subsection{Model verification and validation}\label{sec:validation}
The complexity and great uncertainty of ambulance emergency systems make model verification and validation fundamental for the trust and reliability of the model. All standard verification tools (face validity, model trace, sensitivity analysis) must be adopted to guarantee model validity (see, e.g., \cite{sargent.2013}). Moreover, even if an accurate model validation is always needed when dealing with DES, a model reproducing the ambulance workflow in a region must be subjected to even more thorough validation. 
One of the most critical issue is represented by vehicle movements modelling. In fact, travel times depends on many unpredictable factors and, even if the calibration procedure we introduced in Section~\ref{subsec:travel-times} attenuate such problem, an extensive historical data validation is needed. Several metrics must be derived from available data and compared with the corresponding simulation-derived measures for a significant number of KPIs of interest.

\section{A case study}\label{sec4}
In this section we apply the general purpose DES model described in the previous section to a real case study provided by the {\em Azienda Regionale Emergenza Sanitaria - ARES 118}, \red{an} Italian Regional Emergency \red{Authority}, 
concerning a territory within the Lazio region of Italy, including the medium-size city of Rieti 
along with its province which also consists of sparsely populated areas.
It constitutes a particularly challenging case study due to its demographic and territorial characteristics. It extends over 2,749.16 km$^2$, with an extensive and predominantly mountainous morphology, encompassing 73 municipalities with a total population of about 150,000 inhabitants and a very low average density of around 55 inhabitants per km$^2$ \cite{istat.rieti}. The population is unevenly distributed, with the city of Rieti accounting for roughly one-third of the total, while many peripheral municipalities have fewer than a thousand residents. This configuration complicates the modeling process for EMS also taking into account that travel times are strongly affected by mountainous morphology, winding roads, and variable accessibility. 
\par
As regards the severity code scale, \red{ARES 118 operates within the standard Italian prehospital Triage system, defined by the Italian Ministry of Health, which consists of a color-code to classify the severity of the emergency calls \cite{palma.2014}}. A four colors scale \red{is adopted}: red, yellow,
green and white tags (in decreasing order from the most urgent).
Red and yellow tags are classified as urgent cases, while green and white cases as non-urgent cases.
\par
We implemented the model by using {\sf SimPy~4.1.1},
 an open source Python library for modeling and simulating discrete event systems using process interaction, where system behavior is described by Python generator functions.

\subsection{Ambulance base locations}\label{sec:ambloc}
\begin{figure}[ht]
	\centering
\includegraphics[width=0.75\textwidth]{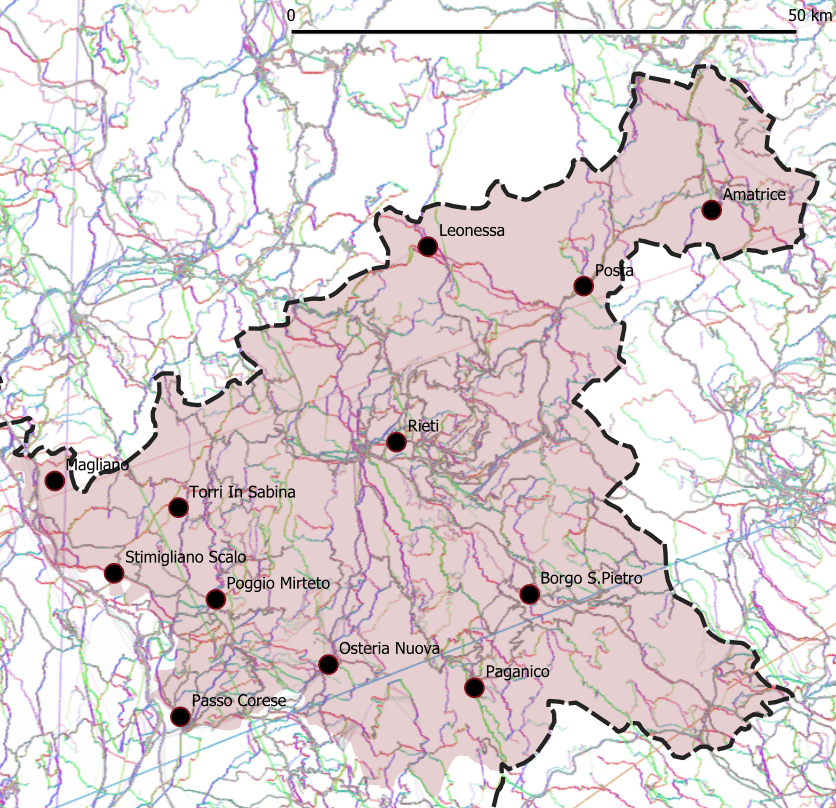}
	\caption{Map of ambulance bases}
	\label{map bases}
\end{figure}

The territory under study currently includes 12 ambulance bases whose locations were not determined using 
some optimization-based/data-driven criterion. 
\red{They were instead established over time using empirical and opportunity criteria, usually in public buildings or other facilities made available for operational use}. 
Figure~\ref{map bases} \red{depicts} their spatial distribution
\red{(their UTM coordinates have been suppressed since they are considered sensitive information)}. 
Each base hosts one ambulance operating on a 24-hour basis (H24), with the exception of the Rieti base, where two H24 ambulances are permanently allocated. In this study, we consider only BLS ambulances. Actually, in the case study area, the EMS system also includes three ALS units, which are physician-staffed emergency vehicles. However, these units cannot transport patients, as they operate as rapid response vehicles providing advanced medical care on scene. Consequently, patient transport always requires the intervention of a BLS ambulance, making BLS units the critical transport resource in the system. For this reason, our analysis focuses on the location and deployment of BLS ambulances. Nevertheless, the proposed framework could be extended in future work to explicitly incorporate ALS units and model their interaction with BLS ambulances.
%

\subsection{Emergency calls database analysis}
We considered all the ambulance dispatches between January 1 and  December 31, 2023 in the territory under study. The raw database consists of 18,611 emergency calls organized in an Excel file where each row represents an accomplished mission, reporting: 
\begin{itemize}	
	\item code associated to the mission;
	\item timestamps of beginning/end of the emergency call (including telephone triage);
	\item physical address and geographical coordinates of patient location (including location type : street, house, office, etc.)
	\item base location of the assigned ambulance;
	\item timestamps of ambulance assignment, departure, arrival on the scene;
	\item patient clinical conditions/hypothesized pathology on the scene;
	\item selected ED;
	\item timestamps of ambulance departure from the scene, arrival at ED, beginning/end of ambulance \red{ramping} (if any), mission end;
	\item outcome.
\end{itemize}
\par
All these data have been used for an accurate input analysis. In particular, by calculating the time differences between pairs of timestamps, we determined all the needed service times indicated in Section~\ref{sec:servicetimes}. Once these service times were collected, a standard best fitting procedure is adopted to determine a theoretical probability distribution. In most cases, due to the high variability of these times, the adopted goodness-of-fit tests lead to reject the null hypothesis at 95\% confidence level. Therefore, empirical distributions are used in these cases.

\subsection{ED locations}\label{sec:edloc}
\bluemax{A map depicting the locations of all the EDs in the Lazio region is reported in Figure \ref{map allhospitals} in the Appendix. Many of them are very far the territory under study and we do not expect that they are actually used for delivering patients.  Moreover,} not all facilities are equipped to manage every type of pathology. Therefore, the dispatcher selects the destination ED by matching the patient clinical condition with the appropriate level of care, following territorial reference rules that link each area to specific hospitals and relying on the regional emergency network, which differentiates between 
general ED capable of managing most emergency cases, but 
without highly specialized services and specialized centers for time-sensitive conditions such as stroke, trauma, and cardiac emergencies (see the description in Section~\ref{sec:pbdescr}). Of course, operational factors such as distance, travel time, and the characteristics of the responding ambulance are also considered, along with real-time information on crowding, boarding, and bed availability, to guarantee the fastest and most appropriate access to definitive care.
%
%
\par
\begin{figure}[htbp]
	\centering
	\includegraphics[width=0.80\textwidth]{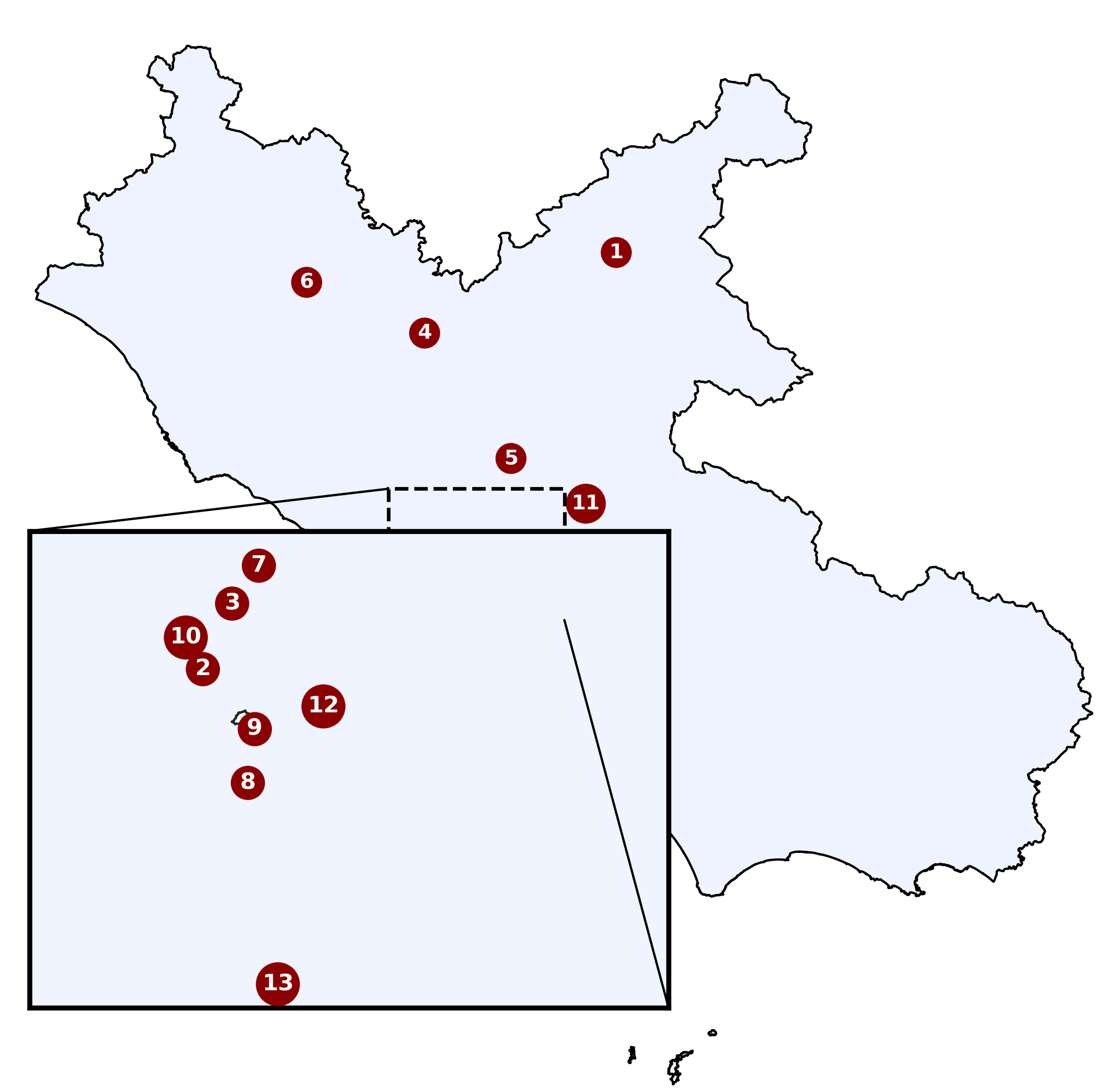}
	\caption{\bluemax{Map of the Lazio Region and the set of potential destination EDs}
		(the numbers in the red circles are in correspondence with the numbered list of Table~\ref{coordinates hospitals}).
		The territory under study (namely, Rieti and its province) is in the North-East part of the region (where the ED~1 is located).}
	\label{map hospitals}
\end{figure}

\bluemax{From available data, by considering all the completed mission, we detected only 13 possible destination EDs that have been actually reached by an ambulance transport from the territory in hand.}
\red{Table~\ref{coordinates hospitals} lists this set of potential destination EDs,} defined according to the criteria established in  Deliberazione n. 869 del 7 dicembre 2023, ``Programmazione della rete ospedaliera 2024-2026 in conformità agli standard previsti nel DM 70/2015'' \cite{Del.869/2023}. A map of the Lazio region of Italy reporting the locations of these EDs  is depicted in Figure~\ref{map hospitals} \red{(also in this case their geographic coordinates have been suppressed)}. Note that the territory under study is located on the North-East part of the Lazio region and it includes only one ED of the network, \bluemax{the number 1 of the list}.  This implies that some of the emergency calls are actually dispatched to EDs outside the territory and, in particular, to EDs located in the province of Rome.
\begin{table}[htbp]
	\centering
	\begin{tabular}{rl}
		\toprule
		1. & Rieti S. Camillo de Lellis – Polo Ospedaliero Unico Integrato \\
		2. &Policlinico A. Gemelli e C.I.C. \\
		3. & Ospedale San Pietro Fatebenefratelli\\
		4. & Ospedale di Civita Castellana\\
		5. & Ospedale SS. Gonfalone  \\
		6. & Viterbo Belcolle – Polo Ospedaliero  \\
		7. & Azienda Ospedaliera Sant’Andrea\\
		8. & A.O. San Camillo–Forlanini  \\
		9. & Ospedale Pediatrico Bambino Gesù  \\
		10. & Presidio Ospedaliero San Filippo Neri  \\
		11. & Ospedale S. Giovanni Evangelista (Tivoli)\\
		12. & Policlinico Umberto I \\
		13. & Policlinico Universitario Campus Bio-Medico  \\
		\bottomrule
	\end{tabular}
	\caption{List of potential EDs}
	\label{coordinates hospitals}
\end{table}

\subsection{Implementation of the model}
\begin{figure}[htbp]
	\centering
	\includegraphics[width=0.9\textwidth]{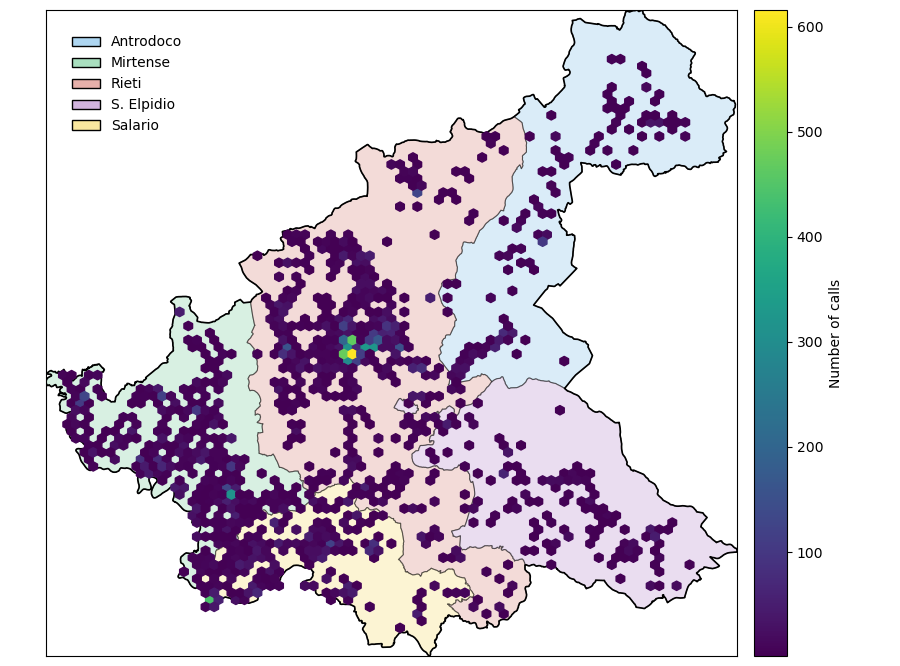}
	\caption{Total number of calls}
	\label{calls on map}
\end{figure}

Now we describe how to apply the general purpose model introduced in Section~\ref{sec:modeldescr}, to this case study. Based on the characteristics of the territory considered, we divided it into 5 \textit{generation zones}
	(\textit{Antrodoco}, \textit{Mirtense}, \textit{Rieti},     
\textit{S. Elpidio}, \textit{Salario}) and, after data cleaning, emergency calls were classified according to these 5 zones and 4 time slots
(00:00-07:00,  07:00-12:00, 12:00-18:00, 18:00-00:00). 
\begin{table}[htbp]
	\centering
	\begin{tabular}{@{}lcccc@{}}
		\toprule
	 & 00:00-07:00 & 07:00-12:00 & 12:00-18:00& 18:00-00:00 \\ 
		\midrule
		\textit{Antrodoco}   & 81 & 161 & 225 & 191 \\
		\textit{Mirtense}    & 348 & 683 & 876 & 692 \\
		\textit{Rieti}       & 782 & 1788 & 2038 & 1641 \\
		\textit{S. Elpidio}  & 90 & 163 & 191 & 140 \\
		\textit{Salario}     & 295 & 513 & 650 & 566 \\
		\bottomrule
	\end{tabular}
	\caption{Number of calls by zone and time slots}
	\label{calls}
\end{table}
The resulting temporal-spatial distribution of dispatches is summarized in Table~\ref{calls}, while the corresponding heatmap illustrating total call intensity across zones 
is reported in Figure~\ref{calls on map}.
It can be clearly observed from this figure a thickening of calls in correspondence of the biggest city in the area (Rieti), while in the remaining zones, rarely populated areas associated to a low number (or even no call) are highlighted.

\subsubsection{Demand points}
\begin{figure}[htbp]
	\centering
	\includegraphics[width=0.85\textwidth]{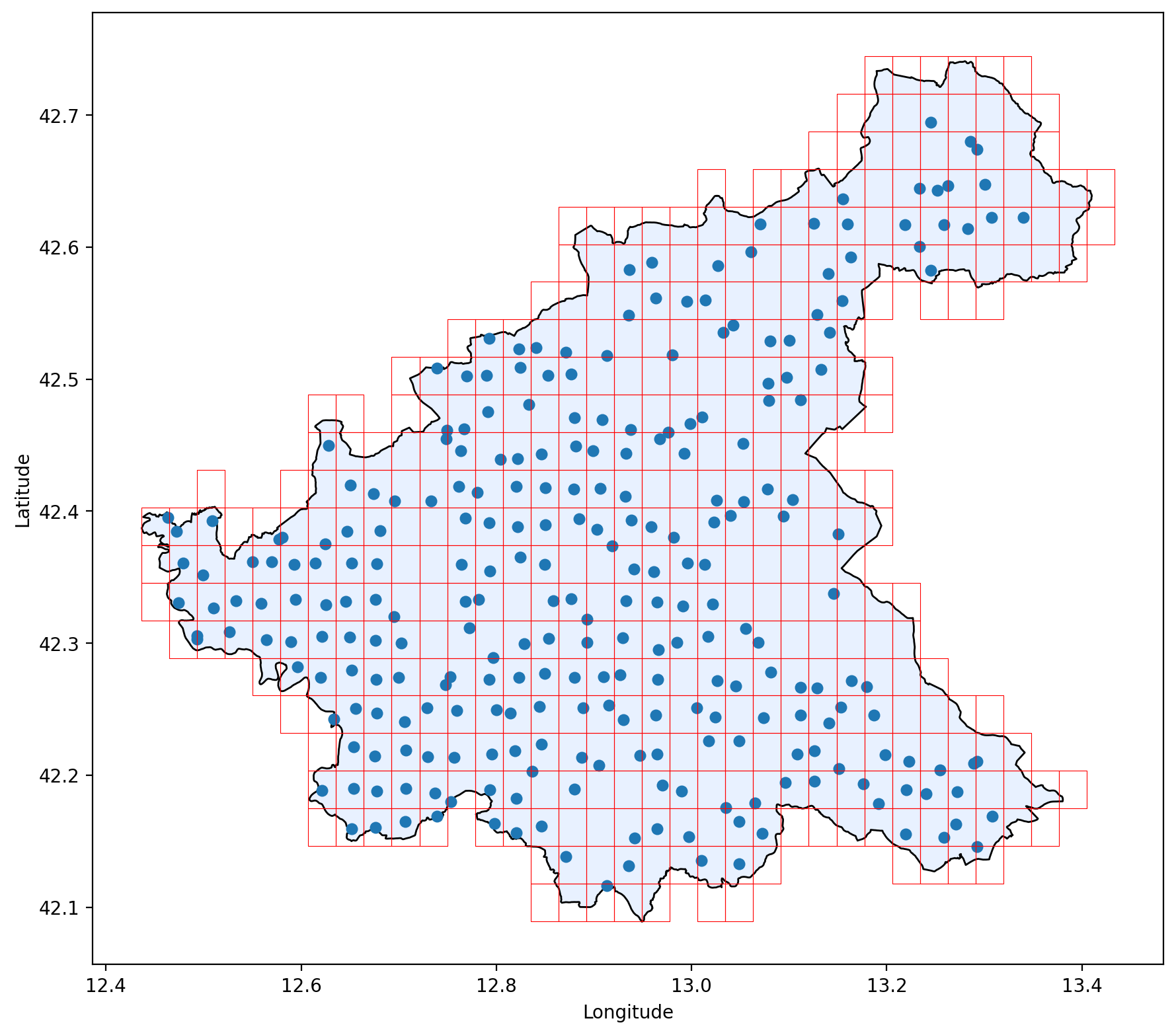}
	\caption{Demand points}
	\label{map demand points}
\end{figure}

To define the demand points, i.e., the locations of the simulated emergency calls, Rieti and its province was partitioned into 446 \textit{call squares}, each with an area of approximately 10 km$^2$. For each square, we selected a specific latitude and longitude coordinate corresponding to the actual call location closest (in Euclidean distance) to the square’s centroid. This approach ensures that each demand point represents a ``feasible origin'',
i.e., a point where emergency call can be really generated.
Squares with no recorded calls were excluded, as no demand was observed from those areas. The resulting grid is shown in Figure \ref{map demand points}.

\subsubsection{Travel times among points of the network}
The model employs the UTM coordinates of the 272 demand points, 12 ambulance bases, and 13 EDs to define the road network and the shortest path between all pairs of point is calculated and stored before the simulation begins (they are usually saved to file and reused for subsequent simulations). The travel times are computed by using the estimates provided by {\sf OpenRouteService} adjusted through correction factors to approximate the actual RTs as described in Section~\ref{subsec:travel-times}. The correction accounted for both the time period (to reflect traffic conditions) and the urgency level of the call (to represent whether the lights-and-sirens travel mode was activated or not). 
Different correction factors are applied to the base-scene segment and to the scene-ED segment, depending on the urgency level. The applied values are reported in Table \ref{correction factors}. The time periods used in the analysis correspond to five operational intervals: weekday peak hours (07:00-09:00 and 17:00-19:00), weekday off-peak hours (06:00-07:00, 09:00-17:00, and 19:00-23:00), weekday night hours (23:00-06:00), weekend and August off-peak hours (06:00-23:00), and weekend and August night hours (23:00-06:00).
\begin{table}[htbp]
	\centering
	\begin{tabular}{@{}lcccc@{}}
		\toprule
		& \multicolumn{2}{c}{\textbf{Non-urgent}} & \multicolumn{2}{c}{\textbf{Urgent}} \\
		\cmidrule(lr){2-3}\cmidrule(lr){4-5}
		\textbf{Time period}
		& Base--scene & Scene--hospital
		& Base--scene & Scene--hospital \\
		\midrule
		Weekday peak hours                
		& \upd{0.983} & \upd{0.737} & \upd{0.897} & \upd{0.723} \\
		
		Weekday off-peak hours              
		& \upd{0.953} & \upd{0.759} & \upd{0.865} & \upd{0.720} \\
		
		Weekday night hours               
		& \upd{1.044} & \upd{0.761} & 0.959 & \upd{0.709} \\
		
		Weekend / August off-peak hours   
		& \upd{0.881} & \upd{0.797} & \upd{0.877} & \upd{0.737} \\
		
		Weekend / August night hours    
		& \upd{1.046} & \upd{0.761} & \upd{0.991} & \upd{0.721} \\
		\bottomrule
	\end{tabular}
		\caption{Correction factors by segment, urgency level, and time interval}\label{correction factors}
\end{table}
\par
As it can be easily observed from this table, for the territory under study, most of the \upd{base--scene} correction factors are close to 1. This should be due to the fact that, on that territory, traffic congestion is not very frequent, hence travel times closely resemble those obtained from the routing service. \upd{The lower scene--hospital factors indicate that, when transporting patients towards the EDs (mostly along main roads), ambulances travel faster than the nominal routing times.}

\subsection{Verification and validation of the model}
The model was first subjected to a verification phase conducted in close collaboration with ARES 118 experts, who examined both the conceptual structure and the simulation logic.
Moreover, standard verification tools like face validity and model trace have been adopted. This phase confirmed the consistency of the modeling assumptions and the coherence of the simulated dynamics with the operational characteristics of the ambulance emergency system in the territory under study.
\par
Model validation was subsequently performed by comparing the values of KPIs derived from historical data with those generated by the simulation, in line with standard validation practices for DES models \cite{sargent.2013}. Validation was articulated through a sequence of complementary analyses, each addressing a different aspect of system behavior.
\par
As regards the simulation runs, we set the length of each replication to 547,200 minutes  (corresponding to twelve months of operations), with a 15-day warm-up period. Moreover, 30 independent replications have been performed for each run.
\par
First, the aggregate number of calls were validated by urgency class. Table \ref{call volume} reports the comparison between historical and simulated annual call counts for urgent (red and yellow tags) and non-urgent (green and white tags) cases. 
\begin{table}[htbp]
	\centering
	\begin{tabular}{lcccccc}
		\toprule
		 &
		$\mu$ &
		\textit{AVG}$_{sim}$ &
		\textit{LB}$_{sim}$ &
		\textit{UB}$_{sim}$ &
		$|\textit{LB}_{sim}-\mu|$ &
		$|\textit{UB}_{sim}-\mu|$ \\
		\midrule
		Urgent      & 10399 & \upd{10390.97} & \upd{10345.19} & \upd{10436.74} & \upd{53.81} & \upd{37.74}  \\
		Non-urgent  & 1715 & \upd{1752.07} & \upd{1736.69} & \upd{1767.44} & \upd{21.69} & \upd{52.44}  \\
		\bottomrule
	\end{tabular}
		\caption{Comparison between historical and simulated \textit{number of calls}}
	\label{call volume}
\end{table}
In particular, $\mu$ denotes the number of call computed from data, \textit{AVG}$_{sim}$ the average simulate number of call and $[\textit{LB}_{sim} , 
	\textit{UB}_{sim}]$ the 95\% confidence interval. 
 Table \ref{call volume} shows that the difference between historical data ($\mu$) and bounds of the confidence interval of simulated number of calls do not exceeded \upd{0.5\%} for urgent calls and \upd{3.1\%} for non-urgent calls. Overall, the differences remains below a tolerance of 5\% for  both urgency classes and
this indicates that discrepancies between real and simulated demand are bounded within a predefined admissible error (e.g., 5\%), supporting the validity of the demand generation process, even if a slight overestimation in the average \upd{number of non-urgent calls} is evidenced.
\par
Second, the model was validated in terms of \textit{coverage}, (percentage of calls for which RT does not exceed a prefix threshold time). 
Table~\ref{tab cov U as is} and Table~\ref{tab cov NU as is} report the coverage values for different values of threshold time (in minutes) obtained from the average of real data and from simulation output for urgent and non-urgent cases, respectively. In Tables~\ref{tab cov U as is} and \ref{tab cov NU as is}, $\mu$ denotes the coverage computed from data, \textit{AVG}$_{sim}$ the average simulated coverage and $[\textit{LB}_{sim} , 
	\textit{UB}_{sim}]$ the 95\% confidence interval. 
 The same results are depicted in Figure~\ref{fig cov U as is} and Figure~\ref{fig cov NU as is}.
\begin{table}[ht]
	\centering
	\begin{tabular}{c c c c c c c}
		\toprule
		\textbf{Threshold} & \(\mu\) & \textit{AVG$_{sim}$} & \textit{LB}$_{sim}$ & \textit{UB}$_{sim}$ &
		$|\textit{LB}_{sim}-\mu|$ &
		$|\textit{UB}_{sim}-\mu|$ \\
		\midrule
		\textit{10} & 16.96\% & \upd{21.89\%} & \upd{21.70\%} & \upd{22.09\%} & \upd{4.74\%} & \upd{5.13\%} \\
		\textit{15} & 42.47\% & \upd{47.00\%} & \upd{46.78\%} & \upd{47.21\%} & \upd{4.32\%} & \upd{4.75\%} \\
		\textit{20} & 61.62\% & \upd{66.72\%} & \upd{66.49\%} & \upd{66.94\%} & \upd{4.87\%} & \upd{5.32\%} \\
		\textit{30} & 86.36\% & \upd{86.00\%} & \upd{85.80\%} & \upd{86.20\%} & \upd{0.56\%} & \upd{0.16\%} \\
		\textit{40} & 95.69\% & \upd{95.08\%} & \upd{94.97\%} & \upd{95.19\%} & \upd{0.72\%} & \upd{0.50\%} \\
		\textit{50} & 98.23\% & \upd{98.23\%} & \upd{98.15\%} & \upd{98.31\%} & \upd{0.08\%} & \upd{0.08\%} \\
		\textit{60} & 99.16\% & \upd{99.14\%} & \upd{99.10\%} & \upd{99.18\%} & \upd{0.06\%} & \upd{0.03\%} \\
		\bottomrule
	\end{tabular}
		\caption{Comparison between historical data and simulated output in terms of \textit{coverage for urgent calls (red and yellow tags)} for different threshold values (in minutes)}
	\label{tab cov U as is}
\end{table}
\begin{table}[ht]
	\centering
	\begin{tabular}{c c c c c c c}
		\toprule
\textbf{Threshold} & \(\mu\) & \textit{AVG$_{sim}$} & \textit{LB}$_{sim}$ & \textit{UB}$_{sim}$ &
$|\textit{LB}_{sim}-\mu|$ &
$|\textit{UB}_{sim}-\mu|$ \\
\midrule
		\textit{10} & 15.70\% & \upd{20.32\%} & \upd{19.92\%} & \upd{20.72\%} & \upd{4.22\%} & \upd{5.02\%} \\
		\textit{15} & 42.18\% & \upd{43.26\%} & \upd{42.78\%} & \upd{43.74\%} & \upd{0.60\%} & \upd{1.56\%} \\
		\textit{20} & 62.99\% & \upd{63.19\%} & \upd{62.78\%} & \upd{63.60\%} & \upd{0.21\%} & \upd{0.61\%} \\
		\textit{30} & 85.30\% & \upd{83.16\%} & \upd{82.75\%} & \upd{83.57\%} & \upd{2.55\%} & \upd{1.73\%} \\
		\textit{40} & 95.20\% & \upd{92.35\%} & \upd{92.03\%} & \upd{92.66\%} & \upd{3.17\%} & \upd{2.54\%} \\
		\textit{50} & 97.94\% & \upd{98.14\%} & \upd{97.99\%} & \upd{98.28\%} & \upd{0.05\%} & \upd{0.34\%} \\
		\textit{60} & 99.07\% & \upd{99.57\%} & \upd{99.51\%} & \upd{99.63\%} & \upd{0.45\%} & \upd{0.57\%} \\
		\bottomrule
	\end{tabular}
		\caption{Comparison between historical data and simulated output in terms of \textit{coverage for non-urgent calls (green and white tags)} for different threshold values (in minutes)}
	\label{tab cov NU as is}
\end{table}
\begin{figure}[htbp]
	\centering
	\begin{tikzpicture}
		\begin{axis}[
			width=0.82\textwidth,
			height=0.45\textwidth,
			xlabel={Response time thresholds (minutes)},
			ylabel={Coverage (\%)},
			xmin=9, xmax=61,
			ymin=0, ymax=100,
			xtick={10,15,20,30,40,50,60},
			xticklabels={10,15,20,30,40,50,60},
			ytick={0,20,40,60,80,100},
			ymajorgrids=true,
			grid=both,
			bar width=7pt,
			legend style={
				at={(0.2,0.95)}, anchor=north,
				legend columns=-1,
				/tikz/every even column/.append style={column sep=8pt},
			},
			legend image code/.code={
				\draw[#1,draw=black,fill=#1] (0cm,-0.1cm) rectangle (0.3cm,0.1cm);
			},
			]
			
			\addplot[
			ybar,
			fill=red!70,
			draw=black,
			error bars/y dir=both,
			error bars/y explicit,
			xshift=-3pt
			]
			coordinates {
				(10,21.89) +- (0.20,0.20)
				(15,47.00) +- (0.21,0.21)
				(20,66.72) +- (0.23,0.23)
				(30,86.00) +- (0.20,0.20)
				(40,95.08) +- (0.11,0.11)
				(50,98.23) +- (0.08,0.08)
				(60,99.14) +- (0.04,0.04)
			};
			\addlegendentry{Simulated}
			
			\addplot[
			ybar,
			fill=blue!65,
			draw=black,
			xshift=+3pt
			]
			coordinates {
				(10,16.96)
				(15,42.47)
				(20,61.62)
				(30,86.36)
				(40,95.69)
				(50,98.23)
				(60,99.16)
			};
  			\addlegendentry{Real}
			
		\end{axis}
	\end{tikzpicture}
	\caption{Comparison between historical data (in blue) and simulated output (in red) in terms of \textit{coverage for urgent calls (red and yellow tags)} for different threshold values}
	\label{fig cov U as is}
\end{figure}
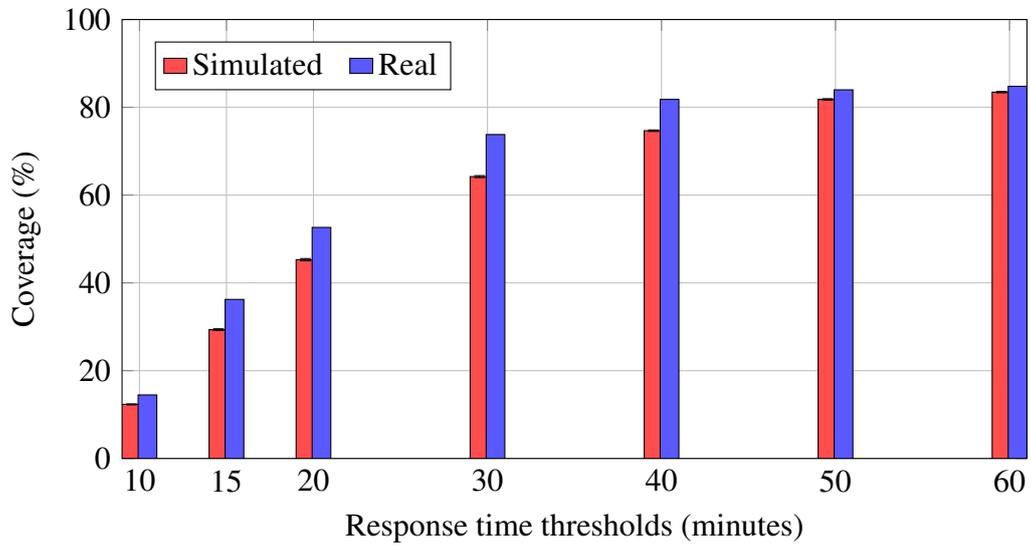
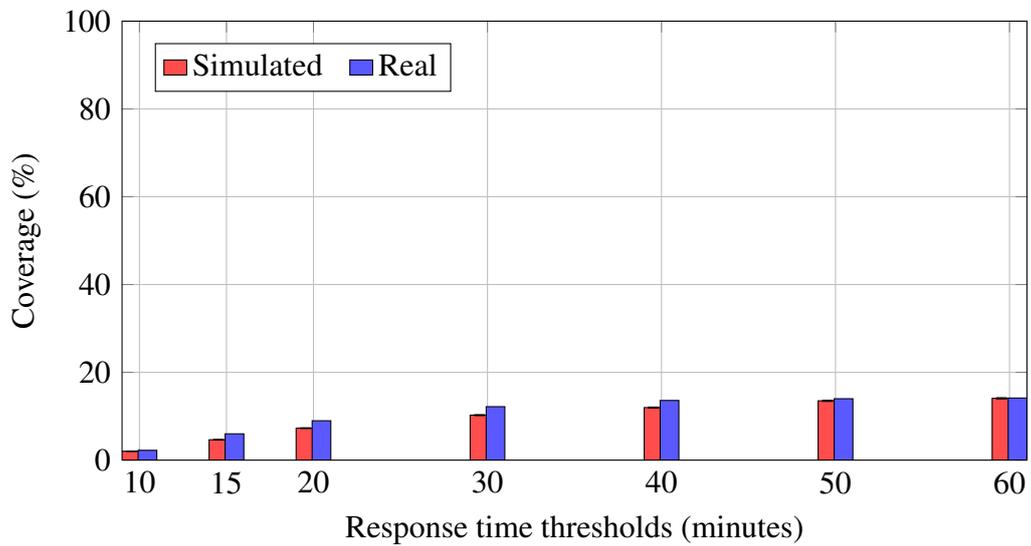
\begin{figure}[htbp]
	\centering
	\begin{tikzpicture}
		\begin{axis}[
			width=0.82\textwidth,
			height=0.45\textwidth,
			xlabel={Response time thresholds (minutes)},
			ylabel={Coverage (\%)},
			xmin=9, xmax=61,
			ymin=0, ymax=100,
			xtick={10,15,20,30,40,50,60},
			xticklabels={10,15,20,30,40,50,60},
			ytick={0, 20,40,60,80,100},
			ymajorgrids=true,
			grid=both,
			bar width=7pt,
			legend style={
				at={(0.2,0.95)}, anchor=north,
				legend columns=-1,
				/tikz/every even column/.append style={column sep=8pt},
			},
			legend image code/.code={
				\draw[#1,draw=black,fill=#1] (0cm,-0.1cm) rectangle (0.3cm,0.1cm);
			},
			]
			
			\addplot[
			ybar,
			fill=red!70,
			draw=black,
			error bars/y dir=both,
			error bars/y explicit,
			xshift=-3pt
			]
			coordinates {
				(10,20.32) +- (0.40,0.40)
				(15,43.26) +- (0.48,0.48)
				(20,63.19) +- (0.41,0.41)
				(30,83.16) +- (0.41,0.41)
				(40,92.35) +- (0.31,0.31)
				(50,98.14) +- (0.15,0.15)
				(60,99.57) +- (0.06,0.06)
			};
			\addlegendentry{Simulated}
			
			\addplot[
			ybar,
			fill=blue!65,
			draw=black,
			xshift=+3pt
			]
			coordinates {
				(10,15.70) 
				(15,42.18) 
				(20,62.99) 
				(30,85.30) 
				(40,95.20) 
				(50,97.94) 
				(60,99.07) 
			};
			\addlegendentry{Real}
			
		\end{axis}
	\end{tikzpicture}
	\caption{Comparison between historical data (in blue) and simulated output (in red) in terms of \textit{coverage for non-urgent calls (green and white tags)} for different threshold values}
	\label{fig cov NU as is}
\end{figure}
\par
These results show that the cumulative RT distributions produced by the simulation closely approximate the real ones, with differences between historical data ($\mu$) and bounds of the confidence interval of simulated coverage that, overall (urgent and non-urgent cases) range from \upd{0.03\%} to \upd{5.32\%} and an average difference of \upd{1.96\%}. These values are consistent with thresholds commonly adopted in the literature on EMS management to consider a DES model sufficiently accurate (see, e.g., \cite{aboueljinane.2014,abohamad.2012}).
\par
Finally, a further validation step concerned \textit{the percentage of emergency calls assigned to each base} (including those calls assigned to ambulances that, after completing the mission on the scene or at the ED, are directly assigned to a new call without actually returning to its own base).
Table~\ref{tab util bases} reports these percentages, comparing historical percentage computed from data with those obtained from the simulation, together with the corresponding confidence intervals. 
In Table~\ref{fig util bases}, $\mu$ denotes the percentage computed from data, \textit{AVG}$_{sim}$ the average percentage from simulation output and $[\textit{LB}_{sim} , 
\textit{UB}_{sim}]$ the 95\% confidence interval. The same results are depicted in Figure~\ref{fig util bases}.
\begin{table}[h!]
	\centering
	\begin{tabular}{lcccccc}
		\toprule
  \textbf{Base} & \(\mu\) & \textit{AVG$_{sim}$} & \textit{LB}$_{sim}$ & \textit{UB}$_{sim}$ &
$|\textit{LB}_{sim}-\mu|$ &
$|\textit{UB}_{sim}-\mu|$ \\
\midrule
		\textit{Rieti} & 42.88\% & \upd{39.93\%} & \upd{39.73\%} & \upd{40.13\%} & \upd{3.16\%} & \upd{2.75\%} \\
		\textit{Osteria Nuova} & 11.47\% & \upd{11.34\%} & \upd{11.25\%} & \upd{11.43\%} & \upd{0.22\%} & \upd{0.04\%} \\
		\textit{Poggio Mirteto} & 8.67\% & \upd{9.40\%} & \upd{9.32\%} & \upd{9.47\%} & \upd{0.64\%} & \upd{0.80\%} \\
		\textit{Torri in Sabina} & 6.57\% & \upd{5.78\%} & \upd{5.70\%} & \upd{5.85\%} & \upd{0.86\%} & \upd{0.72\%} \\
		\textit{Passo Corese} & 6.52\% & \upd{7.82\%} & \upd{7.75\%} & \upd{7.90\%} & \upd{1.23\%} & \upd{1.38\%} \\
		\textit{Borgo S. Pietro} & 6.02\% & \upd{6.61\%} & \upd{6.52\%} & \upd{6.69\%} & \upd{0.50\%} & \upd{0.67\%} \\
		\textit{Stimigliano Scalo} & 5.44\% & \upd{5.87\%} & \upd{5.80\%} & \upd{5.94\%} & \upd{0.36\%} & \upd{0.49\%} \\
		\textit{Posta} & 3.88\% & \upd{3.15\%} & \upd{3.10\%} & \upd{3.20\%} & \upd{0.78\%} & \upd{0.68\%} \\
		\textit{Paganico} & 2.76\% & \upd{3.03\%} & \upd{2.97\%} & \upd{3.09\%} & \upd{0.21\%} & \upd{0.34\%} \\
		\textit{Magliano} & 2.48\% & \upd{2.71\%} & \upd{2.64\%} & \upd{2.77\%} & \upd{0.16\%} & \upd{0.29\%} \\
		\textit{Leonessa} & 2.33\% & \upd{3.52\%} & \upd{3.46\%} & \upd{3.57\%} & \upd{1.13\%} & \upd{1.24\%} \\
		\textit{Amatrice} & 0.97\% & \upd{0.84\%} & \upd{0.82\%} & \upd{0.87\%} & \upd{0.15\%} & \upd{0.10\%} \\
		\bottomrule
	\end{tabular}
		\caption{Comparison between {\em percentage of calls assigned at each base} from real data and from the simulation (together with the corresponding confidence intervals).}
	\label{tab util bases}
\end{table}
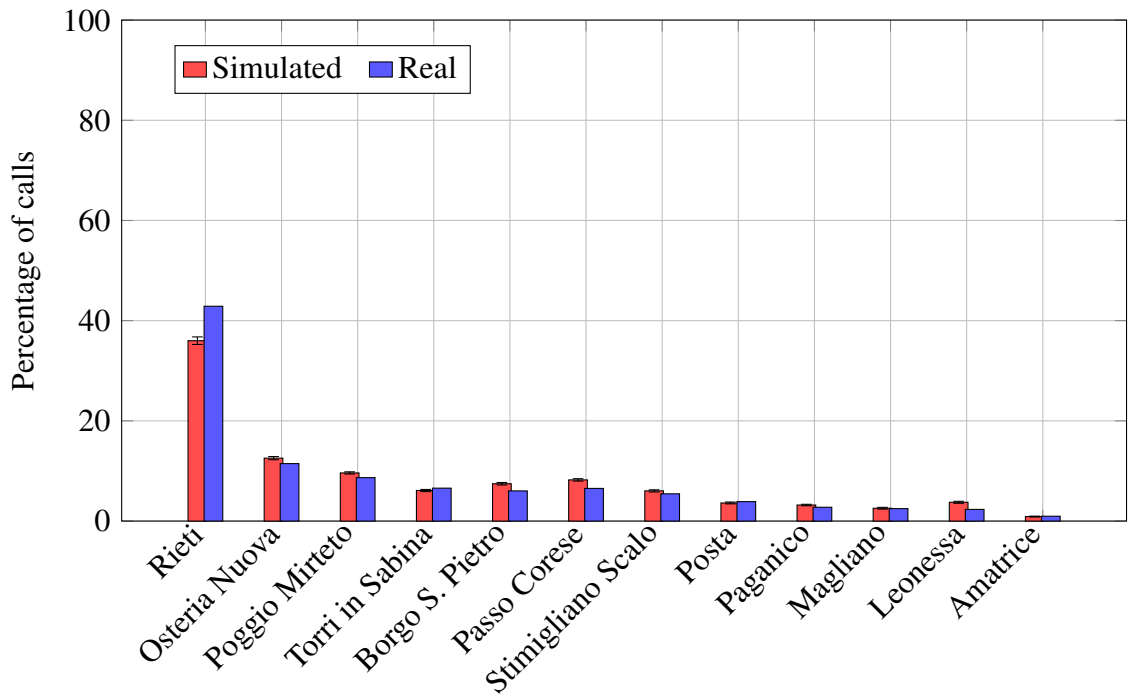
\begin{figure}[htbp]
	\centering
	\begin{tikzpicture}
		\begin{axis}[
			width=0.9\textwidth,
			height=0.5\textwidth,
			ylabel={Percentage of calls},
			ymin=0,
			ymax=100,
			ymajorgrids=true,
			grid=both,
			symbolic x coords={
				Rieti,
				Osteria Nuova,
				Poggio Mirteto,
				Torri in Sabina,
				Borgo S. Pietro,
				Passo Corese,
				Stimigliano Scalo,
				Posta,
				Paganico,
				Magliano,
				Leonessa,
				Amatrice
			},
			xtick=data,
			xticklabel style={rotate=45,anchor=east},
			bar width=7pt,
			legend style={
				at={(0.2,0.95)}, anchor=north,
				legend columns=-1,
				/tikz/every even column/.append style={column sep=8pt},
			},
			legend image code/.code={
				\draw[#1,draw=black,fill=#1] (0cm,-0.1cm) rectangle (0.3cm,0.1cm);
			},
		]
			
		\addplot[
			ybar,
			fill=red!70,
			draw=black,
			error bars/y dir=both,
			error bars/y explicit,
			xshift=-3pt
		]
		coordinates {
			(Rieti,39.93)            += (0,0.20) -= (0,0.20)
			(Osteria Nuova,11.34)    += (0,0.09) -= (0,0.09)
			(Poggio Mirteto,9.40)    += (0,0.08) -= (0,0.08)
			(Torri in Sabina,5.78)   += (0,0.07) -= (0,0.07)
			(Borgo S. Pietro,6.61)   += (0,0.09) -= (0,0.09)
			(Passo Corese,7.82)      += (0,0.08) -= (0,0.08)
			(Stimigliano Scalo,5.87) += (0,0.07) -= (0,0.07)
			(Posta,3.15)             += (0,0.05) -= (0,0.05)
			(Paganico,3.03)          += (0,0.06) -= (0,0.06)
			(Magliano,2.71)          += (0,0.06) -= (0,0.06)
			(Leonessa,3.52)          += (0,0.05) -= (0,0.05)
			(Amatrice,0.84)          += (0,0.03) -= (0,0.03)
		};
		\addlegendentry{Simulated}
			
		\addplot[
			ybar,
			fill=blue!65,
			draw=black,
			xshift=+3pt
		]
		coordinates {
			(Rieti,42.88)
			(Osteria Nuova,11.47)
			(Poggio Mirteto,8.67)
			(Torri in Sabina,6.57)
			(Borgo S. Pietro,6.02)
			(Passo Corese,6.52)
			(Stimigliano Scalo,5.44)
			(Posta,3.88)
			(Paganico,2.76)
			(Magliano,2.48)
			(Leonessa,2.33)
			(Amatrice,0.97)
		};
		\addlegendentry{Real}
			
		\end{axis}
	\end{tikzpicture}
	\caption{Comparison between {\em percentage of calls assigned at each base} from real data (blue) and from the simulation (red).}
	\label{fig util bases}
\end{figure}
\par
For all bases, observed utilizations are either contained within or very close to the simulated confidence interval bounds, 
with differences between historical data ($\mu$) and bounds of the confidence interval of simulated average utilizations
that range from \upd{0.04\%} to \upd{3.16\%} and an average difference of \upd{0.79\%}. This confirms that the model accurately captures both spatial heterogeneity and operational workload patterns across individual ambulance bases.
\par
Overall, the results report in this section, in terms of number of calls, coverage and percentage of calls assigned at each base provide evidence of the validity of the proposed DES model and support its use as an effective decision support system.

\subsection{Design of Experiments and the ``as-is'' status}
Once the DES model has been verified and validated, we performed an accurate design of experiments, so that the model output can be used as representative of the current ``as-is'' status. We confirm the simulation parameters adopted in the validation of the model, namely:
	\begin{itemize}
		\item the length of each replication is set to 380 days;
		\item the length of the warm-up period is set to 15 days;
		\item for each run, 30 independent replications are performed.
	\end{itemize}
\par
With these setting of we obtain one year of simulated system functioning.
As regards the KPI of interest, we focus on the major KPI typically adopted for evaluating an ambulance emergency system, namely the coverage with a threshold value of 20 minutes. That is, we consider the percentage of calls for which ambulance reached the scene within 20 minutes, aggregating urgent calls (red and yellow tags) and non-urgent calls (green and white tags).
\begin{table}[htbp]
	\centering
	\begin{tabular}{cccccc}
		\toprule
		 \multicolumn{3}{c}{\textbf{Urgent}} & \multicolumn{3}{c}{\textbf{Non-urgent}} \\
		\cmidrule(lr){1-3}\cmidrule(rl){4-6}
	\textit{AVG}	& \textit{LB} & \textit{UB} & \textit{AVG}
		& \textit{LB} & \textit{UB} \\
		\cmidrule(lr){1-3}\cmidrule(rl){4-6}
		\upd{66.72\%} & \upd{66.49\%} & \upd{66.94\%} & \upd{63.19\%} & \upd{62.78\%} & \upd{63.60\%} \\
		\bottomrule
\end{tabular}
	\caption{Coverage for urgent and non-urgent calls for 20 minutes threshold value for the \textit{``as-is'' status}. Average value (\textit{AVG}) and 95\% confidence interval $[LB , UB]$}\label{as-is}
\end{table}
These results represent the performance of the current status of the emergency system under study and we will used as benchmark for assessing results obtained from different scenarios. Complete results for different threshold values are reported in the~\ref{secA1}.

\par\bigskip
\subsection{Scenario analysis}\label{sec:analisiscenario}
To show how EMS managers can in practice and fruitfully use the model we proposed, in this section we report the results of a preliminary scenario analysis that allows to evaluate how changes in the ambulance system configuration affect system performance, with particular focus on the coverage KPI. All results are benchmarked versus the ``as-is'' status configuration reported in Section~\ref{sec:ambloc} and in Section~\ref{sec:edloc} whose corresponding coverage values are in Table~\ref{as-is}.
\par
A set of alternative deployment scenarios was defined based on some discussions held with ARES 118 personnel.
Each scenario was simulated over a one year horizon, adopting the same warm-up period and number of replications used for the ``as-is'' status, thus ensuring full comparability of results. Besides H24 ambulances, operating 24 hours a day, H12 ambulances operating only during daytime hours (from 08:00 to 20:00) are also considered. 
\par
\begin{figure}[htbp]
	\centering
\includegraphics[width=0.78\textwidth]{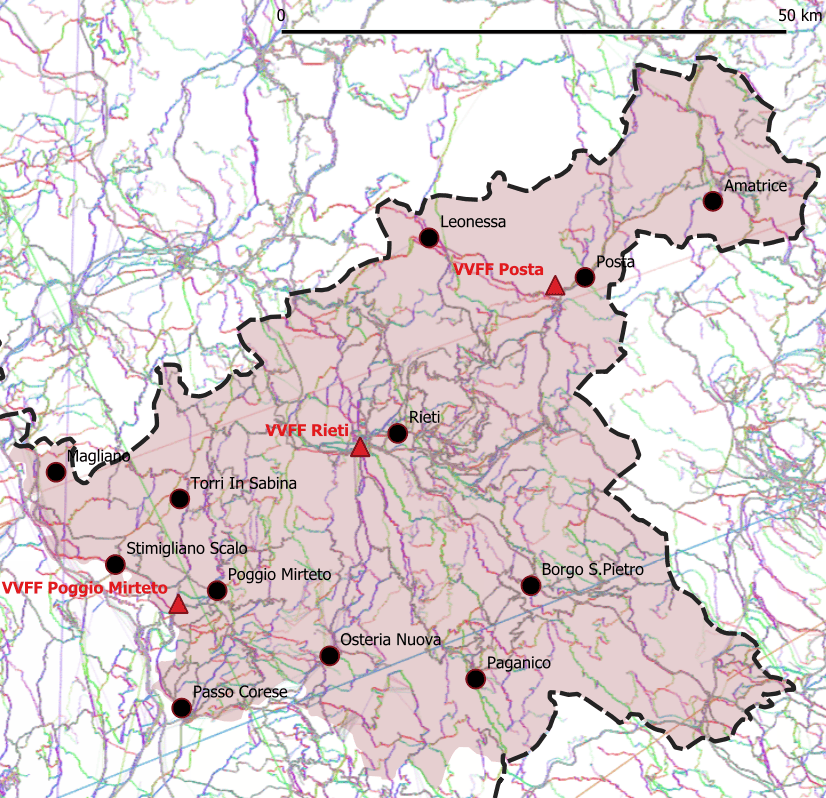}
	\caption{Map of ambulance bases and Fire Stations \bluemax{(the red triangles denoted by VVFF})}
	\label{map basesFS}
\end{figure}

In our experimentation we considered several scenarios. However, for the sake of brevity, we report here only some of the most significant. In particular, \red{as exploratory attempt}, Fire Stations (FSs) were identified as potential candidate sites for the deployment of new ambulance bases. Therefore the three FSs in the territory under study are considered in the scenario analysis as alternative or supplementary locations. 
Figure~\ref{map basesFS} depicts their locations on the map.
\par
The scenarios we explore changes in resource availability and spatial deployment, and are defined as follows:
\par\medskip
\begin{itemize}
	\item \textit{Scenario 1} \textbf{(S1)}: addition of one H12 ambulance at the Rieti base.
	\item \textit{Scenario 2} \textbf{(S2)}: conversion of one H24 ambulance at the Rieti base into an H12 unit.
	\item \textit{Scenario 3} \textbf{(S3)}: addition of one H24 ambulance at the Poggio Mirteto fire station.
	\item \textit{Scenario 4} \textbf{(S4)}: addition of one H24 ambulance at the Posta fire station.
	\item \textit{Scenario 5} \textbf{(S5)}: addition of one H24 ambulance at the Rieti fire station.
	\item \textit{Scenario 6} \textbf{(S6)}: addition of one H24 ambulance at the Poggio Mirteto fire station, combined with the removal of one H24 unit from the closest base.
	\item \textit{Scenario 7} \textbf{(S7)}: addition of one H24 ambulance at the Posta fire station, combined with the removal of one H24 unit from the closest base.
	\item \textit{Scenario 8} \textbf{(S8)}: addition of one H24 ambulance at the Rieti fire station, combined with the removal of one H24 unit from the closest base.
\end{itemize}
\par\medskip
We report in Table~\ref{tab:scenari} a summary of these scenarios along with the total number of the ambulances employed in each of them.
\begin{table}[h]
	\centering
	\begin{tabular}{@{}rlccccccccc@{}}
	\toprule
	&	 &  ``as-is'' &  S1 & S2 &S3&S4&S5&S6&S7&S8\\ 
	\midrule
	1. &	Amatrice       &1 &1  &    1      &  1 & 1 &  1 &   1 &   1 &  1   \\
	2. &	Borgo S.Pietro &1 &1&    1    &  1 & 1 &  1 &   1 &   1 &  1   \\
	3. &	Leonessa       &1 &1&      1     & 1 &  1 &   1 &   1 &   1 &  1        \\
	4. &	Magliano       &1 &1&        1    &  1 & 1 &  1 &   1 &   1 &  1    \\
	5. &	Osteria Nuova  &1 &1&    1     &  1 & 1 &  1 &   1 &   1 &  1    \\
	6. &	Paganico       &1 &1&       1      & 1 &  1 &   1 &   1 &   1 & 1       \\
	7. &	Passo Corese   &1 &1&   1      & 1 &  1 &   1 &   1 &   1 &  1         \\
	8. &	Poggio Mirteto &1 &1&     1     &  1 & 1 &  1 &   0 &   1 &  1    \\
	9. &	Posta          &1 &1&           1    &  1 & 1 &  1 &   1 &   0 &  1     \\
	10. &	Rieti          &2 &2+1$^*$&\upd{1+1$^*$}  &  2 & 2 &  2 &   2 &   2 &  1  \\
	11. &	Stimigliano Scalo &1 &1& 1 &  1 & 1 &  1 &   1 &   1 &  1  \\
	12. &	Torri In Sabina   &1 &1&   1 &  1 & 1 &  1 &   1 &   1 &  1  \\
	13. &	Poggio Mirteto FS &0 &0&0 &  1 &  0 &  0 &   1 &   0 &  0         \\
	14. &	Posta FS          &0 &0&      0&   0& 1  & 0  &  0  &  1  & 0         \\
	15. &	Rieti FS          &0 &0&       0 &   0& 0 &  1 &   0 &   0 &  1       \\
	\midrule
	& & 13 &13+1$^*$&\upd{12+1$^*$} & 14 & 14 & 14 & 13 & 13 & 13 \\
	\bottomrule
\end{tabular}
	\caption{Number of H24 ambulances in each base for different scenarios (1$^*$ denotes an H12 ambulance).
	}\label{tab:scenari}
\end{table}
\par
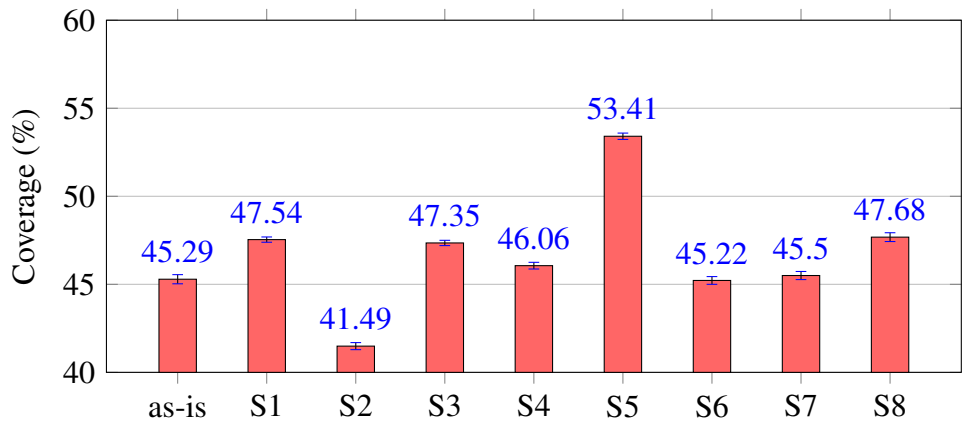
\begin{figure}[htbp]
	\centering
	\begin{tikzpicture}
		\begin{axis}[
			ybar,
			width=0.78\textwidth,
			height=0.38\textwidth,
			ylabel={Coverage (\%)},
			symbolic x coords={as-is,S1,S2,S3,S4,S5,S6,S7,S8},
			xtick=data,
			ymin=40, ymax=80,
			ymajorgrids=true,
			bar width=14pt,
			nodes near coords,
			nodes near coords align={vertical},
			nodes near coords style={yshift=2pt}
			]
			\addplot+[
			fill=red!60,
			draw=black,
			error bars/y dir=both,
			error bars/y explicit,
			]
			coordinates {
				(as-is,66.717) +- (0.226,0.226)
				(S1,70.897)    +- (0.198,0.198)
				(S2,62.889)    +- (0.170,0.170)
				(S3,68.287)    +- (0.227,0.227)
				(S4,67.340)    +- (0.216,0.216)
				(S5,74.345)    +- (0.176,0.176)
				(S6,66.562)    +- (0.179,0.179)
				(S7,66.918)    +- (0.221,0.221)
				(S8,68.730)    +- (0.227,0.227)
			};
		\end{axis}
	\end{tikzpicture}
	\caption{Coverage for 20 minutes threshold value for \textit{urgent calls (red and yellow tags)}. Average and 95\% confidence intervals}
	\label{fig cov20 U}
\end{figure}
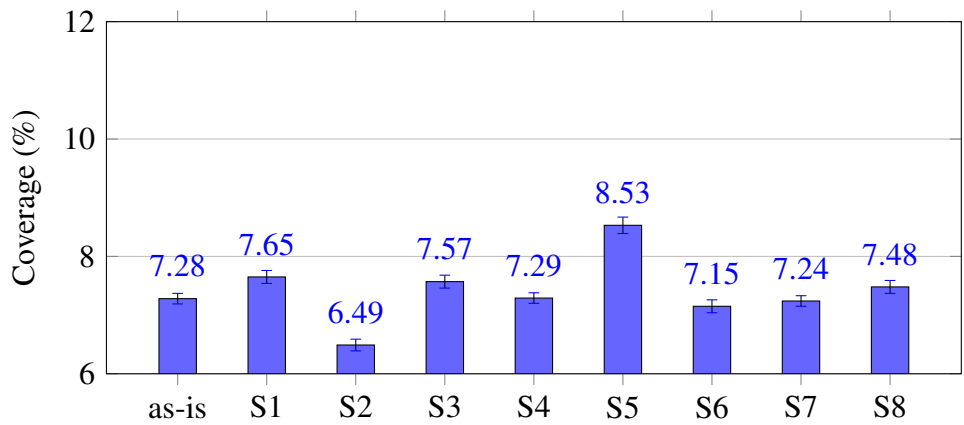
\begin{figure}[htbp]
	\centering
	\begin{tikzpicture}
		\begin{axis}[
			ybar,
			width=0.78\textwidth,
			height=0.38\textwidth,
			ylabel={Coverage (\%)},
			symbolic x coords={as-is,S1,S2,S3,S4,S5,S6,S7,S8},
			xtick=data,
			ymin=40, ymax=80,
			ymajorgrids=true,
			bar width=14pt,
			nodes near coords,
			nodes near coords align={vertical},
			nodes near coords style={yshift=4pt}
			]
			\addplot+[
			fill=blue!60,
			draw=black,
			error bars/y dir=both,
			error bars/y explicit,
			]
			coordinates {
				(as-is,63.189) +- (0.412,0.412)
				(S1,67.535)    +- (0.433,0.433)
				(S2,59.891)    +- (0.463,0.463)
				(S3,64.720)    +- (0.532,0.532)
				(S4,64.349)    +- (0.480,0.480)
				(S5,71.398)    +- (0.536,0.536)
				(S6,63.309)    +- (0.440,0.440)
				(S7,64.069)    +- (0.641,0.641)
				(S8,65.663)    +- (0.404,0.404)
			};
		\end{axis}
	\end{tikzpicture}
	\caption{Coverage for 20 minutes threshold value for \textit{non-urgent calls (green and white tags)}. Average and 95\% confidence intervals}
	\label{fig cov20 NU}
\end{figure}

Note that Scenarios 1, 3, 4, and 5 required an increase (1 ambulance added) of the overall number of ambulances employed. Scenarios 6, 7, 8 correspond to a movement of an H24 ambulance from a base to the closest Fire Station. Scenario~2 only considers a conversion of one of the two H24 ambulances currently assigned to Rieti base to an H12 operating unit. These scenarios represent simple changes with respect to the ``as-is'' configuration. However, it is worth noting that any different ambulance bases location, adding and/or removing bases and/or vehicles, and/or different ambulance deployments among the bases can be easily considered in our model. Figure~\ref{fig cov20 U} and Figure~\ref{fig cov20 NU} report average coverage
(for 20 minutes threshold value) and 95\% confidence intervals for urgent calls and non-urgent calls, respectively. Results for the ``as-is'' status are also reported in the figures for comparison. The complete results
including comparisons between the ``as-is'' status and all the considered scenarios and for different threshold values are reported 
in Table~\ref{tab cov s01234} and Table~\ref{tab cov s5678} in the \ref{secA1}.

\par
Figures~\ref{fig cov20 U} and \ref{fig cov20 NU} provide a descriptive comparison in terms of average coverage, showing that, for 20 minute threshold, Scenario~5 yields the largest and most consistent gains. However, to be thorough, we now formally asses statistical significance of performance differences through the standard methodology given by the paired t-confidence intervals computed across replications.

For each replication and for each combination of RT threshold (10, 15, 20, 30, 40, 50, and 60 minutes) and urgency level (urgent and non-urgent), the coverage difference between the ``as-is'' configuration and each alternative scenario is computed. Table~\ref{tab:paired_t_20min} reports the paired t-test results for the 20 minute threshold, between the ``as-is'' status and the considered scenarios, distinguishing urgent and non-urgent calls. \red{Note that since we do not assume that mean differences are normally distributed, we rely on the central limit theorem and, as commonly adopted, this implies that the coverage probability will be close to the required confidence as sample dimension grows.}
\begin{table}[htbp]
	\centering
	\begin{tabular}{lcccc}
		\toprule
		\textit{Scenario} & \textit{Urgency} & \textit{Mean diff.} & \textit{95\% CI} & \textit{Result} \\
		\midrule
		\textbf{S1} & Non-urgent & \upd{+4.35} & \upd{(+3.75 , +4.94)} & Significant improvement \\
		\textbf{S1} & Urgent     & \upd{+4.18} & \upd{(+3.90 , +4.46)} & Significant improvement \\
		\hline
		\textbf{S2} & Non-urgent & \upd{-3.30} & \upd{(-3.96 , -2.64)} & Significant worsening \\
		\textbf{S2} & Urgent     & \upd{-3.83} & \upd{(-4.07 , -3.59)} & Significant worsening \\
		\hline
		\textbf{S3} & Non-urgent & \upd{+1.53} & \upd{(+0.85 , +2.21)} & Significant improvement \\
		\textbf{S3} & Urgent     & \upd{+1.57} & \upd{(+1.30 , +1.84)} & Significant improvement \\
		\hline
		\textbf{S4} & Non-urgent & \upd{+1.16} & \upd{(+0.62 , +1.70)} & \upd{Significant improvement} \\
		\textbf{S4} & Urgent     & \upd{+0.62} & \upd{(+0.31 , +0.93)} & Significant improvement \\
		\hline
		\textbf{S5} & Non-urgent & \upd{+8.21} & \upd{(+7.53 , +8.89)} & Significant improvement \\
		\textbf{S5} & Urgent     & \upd{+7.63} & \upd{(+7.31 , +7.94)} & Significant improvement \\
		\hline
		\textbf{S6} & Non-urgent & \upd{+0.12} & \upd{(-0.52 , +0.75)} & \upd{Not significant} \\
		\textbf{S6} & Urgent     & \upd{-0.16} & \upd{(-0.39 , +0.08)} & \upd{Not significant} \\
		\hline
		\textbf{S7} & Non-urgent & \upd{+0.88} & \upd{(+0.18 , +1.58)} & \upd{Significant improvement} \\
		\textbf{S7} & Urgent     & \upd{+0.20} & \upd{(-0.10 , +0.50)} & Not significant \\
		\hline
		\textbf{S8} & Non-urgent & \upd{+2.47} & \upd{(+1.85 , +3.10)} & Significant improvement \\
		\textbf{S8} & Urgent     & \upd{+2.01} & \upd{(+1.71 , +2.32)} & Significant improvement \\
		\bottomrule
	\end{tabular}
		\caption{Paired t-test results for 20 minute threshold between the ``as-is'' status and different scenarios}
	\label{tab:paired_t_20min}
\end{table} 
A 95\% paired t-confidence interval for the mean differences is reported and a scenario is considered to significantly improve performance when the entire confidence interval lies above zero; conversely, it significantly worsens performance when the confidence interval lies entirely below zero.
The same comparison is performed for each value of threshold values, resulting in overall 14 combinations (7 thresholds values $\times$  2 urgency classes). For the sake of brevity, we do not report all these results, but in Table~\ref{tab:paired_all_thresholds} we summarize, for each scenario, the numebr of times it exhibits a statistically significant improvement or worsening.
\begin{table}[htbp]
	\centering
	\begin{tabular}{lccc}
		\toprule
		 & \textit{Significant} & \textit{Significant} &  \\
		\textit{Scenario} & \textit{improvements} & \textit{worsenings} & \textit{Interpretation} \\
		\midrule
		\textbf{S5} & 14 & 0 & Dominant (robust-wide improvement) \\
		\textbf{S1} & \upd{14} & 0 & \upd{Dominant (robust-wide improvement)} \\
		\textbf{S3} & \upd{13} & 0 & Strong improvement \\
		\textbf{S4} & 11 & 0 & Strong improvement \\
		\textbf{S8} & \upd{9} & 2 & Mixed (trade-offs) \\
		\textbf{S7} & \upd{2} & \upd{0} & \upd{Essentially neutral} \\
		\textbf{S6} & 0 & \upd{4} & Mild negative \\
		\textbf{S2} & 0 & \upd{13} & Systematically worsening \\
		\bottomrule
	\end{tabular}
		\caption{Paired t-test summary across all RT thresholds and urgency levels}
	\label{tab:paired_all_thresholds}
\end{table}

Scenario 5, corresponding to the addition of one H24 ambulance at the Rieti fire station, exhibits statistical dominance \upd{in terms of effect size}. It produces significant improvements in all threshold-urgency combinations\upd{, as Scenario~1 does, but with gains about twice as large}.

Scenarios 1, 3 and 4 also generate systematic and statistically significant improvements across most thresholds, although with smaller effect sizes compared to Scenario 5. 

Scenario 2 systematically deteriorates system performance. The paired t-tests indicate statistically significant worsening across all thresholds and urgency classes\upd{, with the only exception of non-urgent calls at 10 minutes}. This confirms that reducing full-time ambulance availability in a high-demand node structurally undermines coverage.

Redistribution scenarios without an increase in net capacity show heterogeneous behavior. Scenario 6 presents several statistically significant deteriorations and no consistent gains. Scenario~7 is largely statistically neutral, with differences indistinguishable from the baseline. Scenario 8 displays mixed effects, combining improvements and deteriorations depending on the threshold, indicating spatial trade-offs rather than uniform system enhancement.
\par
For low threshold values (10-15 minutes), capacity expansion scenarios already produce statistically significant improvements, particularly for urgent demand. For intermediate threshold values (20-30 minutes), differences become more pronounced and robust. For higher thresholds values (50-60 minutes), coverage levels approach saturation, reducing absolute differences; nevertheless, Scenario 5 maintains statistical significance in most cases, demonstrating structural impact beyond marginal adjustments. \upd{The effects of redistribution strategies are concentrated on low threshold values} and tend to become neutral for higher threshold values, reinforcing the conclusion that, in the considered case, spatial reshuffling alone cannot substitute net capacity expansion.

\par
The combined descriptive and inferential evidence leads to a consistent managerial conclusion: structural capacity reinforcement at a central high-demand location produces robust, system-wide improvements across all RT thresholds. Reducing full-time availability leads to systematic deterioration, while redistribution without additional resources yields limited or unstable benefits.

Therefore, from both a statistical and operational perspective, Scenario 5 represents the most effective deployment strategy for improving EMS ambulance coverage in the considered territory, namely Rieti and its province.

\section{Conclusions and future research}\label{sec5}
This paper presented a DES model designed to realistically reproduce the operational dynamics of a regional EMS system and to support data-driven decisions on ambulance deployment. The proposed model integrates a detailed representation of the full EMS workflow, including stochastic call generation with spatial and temporal heterogeneity, calibrated travel times based on historical observations, differentiated service-time distributions by urgency level, hospital referral mechanisms consistent with hub-and-spoke clinical networks, and \red{ramping} phenomena at EDs. In addition, the model relaxes several simplifying assumptions commonly adopted in the literature, most notably the assumption that ambulances systematically return to their home base after completing a mission, thereby achieving a higher degree of operational realism.

The model was validated using a comprehensive real-world case study provided by ARES 118, focusing on a territory in the Lazio region of Italy:
Rieti and its province. It is a territory characterized by strong heterogeneity in population density, demand intensity, and geographical accessibility. Validation results show a close agreement between simulated and historical data in terms of annual call volumes, response-time coverage across multiple thresholds, and ambulance utilization at the base level. Deviations between simulated and observed indicators remain within bounds commonly accepted in the simulation literature, confirming that the proposed DES model provides an accurate and reliable representation of the underlying EMS system and can be confidently employed as a decision support tool.
\par
Beyond reproducing the current configuration, the framework enables a systematic evaluation of alternative deployment scenarios. The scenario analysis highlights that increasing ambulance availability in high-demand and strategically central locations yields the most significant and robust improvements in system performance. In particular, the addition of a full-time (H24) ambulance at the Rieti fire station produces the largest gains in coverage for both urgent and non-urgent calls, reflecting the strong concentration of population and call volume in that area. The analysis also shows that partial-time reinforcement through H12 ambulances can be effective when deployed in zones with high daytime demand, although their impact is intrinsically limited by reduced temporal availability. 
Conversely, redistribution strategies without net increases in resources produce heterogeneous effects: moving a unit toward the Rieti fire station still yields relevant improvements, whereas other relocations are either mostly neutral or associated with performance deterioration.

These results confirm that deployment decisions should explicitly account for demand concentration, temporal availability of resources, and the interaction between spatial positioning and stochastic system dynamics. More generally, the study demonstrates the value of simulation-based approaches in overcoming the limitations of static or purely analytical models, which may fail to capture ambulance unavailability, \red{ramping} at ED effects, and dynamic dispatching behavior.

From a methodological perspective, the proposed framework represents a solid foundation for more advanced decision support systems. Future research directions include the integration of demand forecasting models to enable proactive and adaptive deployment (and redeployment) policies, and the embedding of the DES within a simulation-optimization framework to systematically search for optimal fleet compositions and base configurations under multiple and possibly conflicting objectives. 
\par
Overall, the proposed DES framework provides a flexible and reusable tool for improving EMS planning and resource allocation. One of its key strengths is its full scalability and adaptability, enabling it to effectively accommodate varying system sizes, demand levels, and operational complexities. It also paves the way for the development of adaptive, data-driven EMS management systems. Providing EMS managers with such a decision-support tool would yield significant benefits in terms of both efficiency and equity across the entire EMS.

\section{Acknowledgement}
The research activity of G. Riccardi is
funded by the European Union – Next Generation EU, Mission 4, Component 1
CUP: B53C23002330006.

\section{\upd{Code and data availability}}
\upd{The source code of the simulation model and of the parameter estimation procedure is publicly available at \url{https://github.com/ems-sapienza/ems-simulator-rieti}. The data used in this study were provided by ARES 118 and are not publicly available.}

\appendix
\setcounter{table}{0}
\setcounter{figure}{0}
	
	\section{Coverage comparison between the ``as-is'' status and the considered scenarios}\label{secA1}
In this Appendix we report the complete detailed results of the scenario analysis described in Section~\ref{sec:analisiscenario}. 
In particular, Table~\ref{tab cov s01234} reports the coverage obtained for scenario \textbf{S1}, \textbf{S2}, \textbf{S3}, \textbf{S4} along with the ''as-is'' status (denoted as scenario \textbf{S0}) for different values of the threshold; Table~\ref{tab cov s5678} reports the coverage obtained for scenario \textbf{S5}, \textbf{S6}, \textbf{S7}, \textbf{S8}.

\begin{sidewaystable}	
	\centering
	\begin{tabular}{c|rrr|rrr|rrr|rrr|rrr|}
		\toprule
		&\multicolumn{3}{c|}{\textbf{As-is}} &\multicolumn{3}{c|}{\textbf{S1}}
		&\multicolumn{3}{c|}{\textbf{S2}}
		&\multicolumn{3}{c|}{\textbf{S3}}
		&\multicolumn{3}{c|}{\textbf{S4}}
		\\
		\midrule
		\textbf{Thr.} & \textit{AVG}$_0$ &\textit{LB}$_0$ & \textit{UB}$_0$&  
		\textit{AVG}$_1$ & \textit{LB}$_1$ & \textit{UB}$_1$ &
		\textit{AVG}$_2$ & \textit{LB}$_2$ & \textit{UB}$_2$  &
		\textit{AVG}$_3$ & \textit{LB}$_3$ & \textit{UB}$_3$  &
		\textit{AVG}$_4$ & \textit{LB}$_4$ & \textit{UB}$_4$
		\\
		\midrule
		\textit{min.}	&
		\multicolumn{15}{|c|}{\textbf{Non-urgent}} \\
		\midrule
		10 & \upd{20.32} & \upd{19.92} & \upd{20.72} & \upd{21.90} & \upd{21.53} & \upd{22.26} & \upd{19.85} & \upd{19.51} & \upd{20.18} & \upd{21.64} & \upd{21.37} & \upd{21.92} & \upd{21.09} & \upd{20.73} & \upd{21.45} \\
		15 & \upd{43.26} & \upd{42.78} & \upd{43.74} & \upd{46.40} & \upd{45.98} & \upd{46.81} & \upd{41.12} & \upd{40.67} & \upd{41.57} & \upd{44.81} & \upd{44.39} & \upd{45.22} & \upd{44.39} & \upd{43.90} & \upd{44.89} \\
		20 & \upd{63.19} & \upd{62.78} & \upd{63.60} & \upd{67.53} & \upd{67.10} & \upd{67.97} & \upd{59.89} & \upd{59.43} & \upd{60.35} & \upd{64.72} & \upd{64.19} & \upd{65.25} & \upd{64.35} & \upd{63.87} & \upd{64.83} \\
		30 & \upd{83.16} & \upd{82.75} & \upd{83.57} & \upd{88.12} & \upd{87.81} & \upd{88.44} & \upd{79.23} & \upd{78.85} & \upd{79.61} & \upd{84.12} & \upd{83.80} & \upd{84.44} & \upd{84.42} & \upd{84.02} & \upd{84.81} \\
		40 & \upd{92.35} & \upd{92.03} & \upd{92.66} & \upd{95.32} & \upd{95.09} & \upd{95.55} & \upd{89.86} & \upd{89.54} & \upd{90.18} & \upd{92.82} & \upd{92.61} & \upd{93.03} & \upd{93.09} & \upd{92.87} & \upd{93.32} \\
		50 & \upd{98.14} & \upd{97.99} & \upd{98.28} & \upd{98.71} & \upd{98.61} & \upd{98.81} & \upd{97.66} & \upd{97.52} & \upd{97.80} & \upd{98.33} & \upd{98.20} & \upd{98.47} & \upd{98.56} & \upd{98.43} & \upd{98.69} \\
		60 & \upd{99.57} & \upd{99.51} & \upd{99.63} & \upd{99.71} & \upd{99.67} & \upd{99.75} & \upd{99.44} & \upd{99.37} & \upd{99.50} & \upd{99.59} & \upd{99.53} & \upd{99.66} & \upd{99.64} & \upd{99.58} & \upd{99.69} \\
		\midrule
		\textit{min.}& \multicolumn{15}{c|}{\textbf{Urgent}} \\
		\midrule
		10 & \upd{21.89} & \upd{21.70} & \upd{22.09} & \upd{23.05} & \upd{22.90} & \upd{23.19} & \upd{21.02} & \upd{20.84} & \upd{21.19} & \upd{22.95} & \upd{22.77} & \upd{23.14} & \upd{22.16} & \upd{21.98} & \upd{22.33} \\
		15 & \upd{47.00} & \upd{46.78} & \upd{47.21} & \upd{50.21} & \upd{50.04} & \upd{50.38} & \upd{44.44} & \upd{44.24} & \upd{44.63} & \upd{48.52} & \upd{48.30} & \upd{48.74} & \upd{47.55} & \upd{47.30} & \upd{47.80} \\
		20 & \upd{66.72} & \upd{66.49} & \upd{66.94} & \upd{70.90} & \upd{70.70} & \upd{71.09} & \upd{62.89} & \upd{62.72} & \upd{63.06} & \upd{68.29} & \upd{68.06} & \upd{68.51} & \upd{67.34} & \upd{67.12} & \upd{67.56} \\
		30 & \upd{86.00} & \upd{85.80} & \upd{86.20} & \upd{90.21} & \upd{90.08} & \upd{90.33} & \upd{81.62} & \upd{81.42} & \upd{81.82} & \upd{86.71} & \upd{86.55} & \upd{86.87} & \upd{86.58} & \upd{86.39} & \upd{86.77} \\
		40 & \upd{95.08} & \upd{94.97} & \upd{95.19} & \upd{96.52} & \upd{96.44} & \upd{96.59} & \upd{93.03} & \upd{92.89} & \upd{93.17} & \upd{95.49} & \upd{95.38} & \upd{95.60} & \upd{95.24} & \upd{95.12} & \upd{95.35} \\
		50 & \upd{98.23} & \upd{98.15} & \upd{98.31} & \upd{98.75} & \upd{98.70} & \upd{98.80} & \upd{97.76} & \upd{97.69} & \upd{97.83} & \upd{98.61} & \upd{98.55} & \upd{98.67} & \upd{98.34} & \upd{98.28} & \upd{98.40} \\
		60 & \upd{99.14} & \upd{99.10} & \upd{99.18} & \upd{99.42} & \upd{99.40} & \upd{99.45} & \upd{98.87} & \upd{98.82} & \upd{98.92} & \upd{99.42} & \upd{99.38} & \upd{99.45} & \upd{99.18} & \upd{99.14} & \upd{99.22} \\
		\bottomrule
	\end{tabular}
	\caption{Average coverage (\textit{AVG$_i$}) and 95\% confidence interval $[LB_i , UB_i]$. ``As-is'' status (index $i=0$), Scenario~$1$-Scenario~$4$, for different values of threshold (\textbf{Thr.}) in minutes}
	\label{tab cov s01234}
\end{sidewaystable}
\begin{sidewaystable}	
	\centering
	\begin{tabular}{c|rrr|rrr|rrr|rrr|}
		\toprule
		&\multicolumn{3}{c|}{\textbf{S5}} &\multicolumn{3}{c|}{\textbf{S6}}
		&\multicolumn{3}{c|}{\textbf{S7}}
		&\multicolumn{3}{c|}{\textbf{S8}}
		\\
		\midrule
		\textbf{Thr.} & \textit{AVG}$_5$ &\textit{LB}$_5$ & \textit{UB}$_5$&  
		\textit{AVG}$_6$ & \textit{LB}$_6$ & \textit{UB}$_6$ &
		\textit{AVG}$_7$ & \textit{LB}$_7$ & \textit{UB}$_7$  &
		\textit{AVG}$_8$ & \textit{LB}$_8$ & \textit{UB}$_8$  
		\\
		\midrule
		\textit{min.}	&
		\multicolumn{12}{|c|}{\textbf{Non-urgent}} \\
		\midrule
		10 & \upd{27.96} & \upd{27.49} & \upd{28.42} & \upd{19.40} & \upd{19.01} & \upd{19.80} & \upd{20.70} & \upd{20.30} & \upd{21.10} & \upd{25.63} & \upd{25.33} & \upd{25.93} \\
		15 & \upd{51.84} & \upd{51.23} & \upd{52.45} & \upd{42.92} & \upd{42.52} & \upd{43.31} & \upd{44.05} & \upd{43.40} & \upd{44.70} & \upd{47.46} & \upd{47.04} & \upd{47.88} \\
		20 & \upd{71.40} & \upd{70.86} & \upd{71.93} & \upd{63.31} & \upd{62.87} & \upd{63.75} & \upd{64.07} & \upd{63.43} & \upd{64.71} & \upd{65.66} & \upd{65.26} & \upd{66.07} \\
		30 & \upd{90.41} & \upd{90.06} & \upd{90.75} & \upd{83.27} & \upd{82.96} & \upd{83.57} & \upd{83.55} & \upd{83.08} & \upd{84.02} & \upd{83.44} & \upd{83.09} & \upd{83.79} \\
		40 & \upd{96.65} & \upd{96.44} & \upd{96.87} & \upd{92.39} & \upd{92.16} & \upd{92.63} & \upd{92.61} & \upd{92.34} & \upd{92.89} & \upd{92.52} & \upd{92.25} & \upd{92.80} \\
		50 & \upd{99.17} & \upd{99.08} & \upd{99.27} & \upd{98.18} & \upd{98.05} & \upd{98.32} & \upd{98.20} & \upd{98.05} & \upd{98.35} & \upd{98.38} & \upd{98.25} & \upd{98.52} \\
		60 & \upd{99.78} & \upd{99.73} & \upd{99.83} & \upd{99.55} & \upd{99.48} & \upd{99.62} & \upd{99.62} & \upd{99.55} & \upd{99.69} & \upd{99.68} & \upd{99.62} & \upd{99.74} \\
		\midrule
		\textit{min.}& \multicolumn{12}{c|}{\textbf{Urgent}} \\
		\midrule
		10 & \upd{29.69} & \upd{29.48} & \upd{29.89} & \upd{20.36} & \upd{20.20} & \upd{20.52} & \upd{21.86} & \upd{21.68} & \upd{22.04} & \upd{27.13} & \upd{26.99} & \upd{27.28} \\
		15 & \upd{55.04} & \upd{54.84} & \upd{55.25} & \upd{46.30} & \upd{46.08} & \upd{46.52} & \upd{46.99} & \upd{46.77} & \upd{47.22} & \upd{50.49} & \upd{50.28} & \upd{50.70} \\
		20 & \upd{74.35} & \upd{74.17} & \upd{74.52} & \upd{66.56} & \upd{66.38} & \upd{66.74} & \upd{66.92} & \upd{66.70} & \upd{67.14} & \upd{68.73} & \upd{68.50} & \upd{68.96} \\
		30 & \upd{92.16} & \upd{92.04} & \upd{92.28} & \upd{85.88} & \upd{85.77} & \upd{86.00} & \upd{86.22} & \upd{86.06} & \upd{86.38} & \upd{86.63} & \upd{86.47} & \upd{86.79} \\
		40 & \upd{97.32} & \upd{97.24} & \upd{97.39} & \upd{94.92} & \upd{94.83} & \upd{95.01} & \upd{95.08} & \upd{94.98} & \upd{95.18} & \upd{95.06} & \upd{94.95} & \upd{95.18} \\
		50 & \upd{98.82} & \upd{98.77} & \upd{98.86} & \upd{98.28} & \upd{98.22} & \upd{98.33} & \upd{98.29} & \upd{98.23} & \upd{98.34} & \upd{97.89} & \upd{97.81} & \upd{97.96} \\
		60 & \upd{99.37} & \upd{99.34} & \upd{99.41} & \upd{99.14} & \upd{99.10} & \upd{99.17} & \upd{99.14} & \upd{99.10} & \upd{99.19} & \upd{98.83} & \upd{98.78} & \upd{98.88} \\
		\bottomrule
	\end{tabular}
	\caption{Average coverage (\textit{AVG$_i$}) and 95\% confidence interval $[LB_i , UB_i]$. Scenario~$5$-Scenario~$8$, for different values of threshold (\textbf{Thr.}) in minutes}
	\label{tab cov s5678}
\end{sidewaystable}

\newpage
\begin{figure}[htbp]
	\centering
	\includegraphics[width=0.80\textwidth]{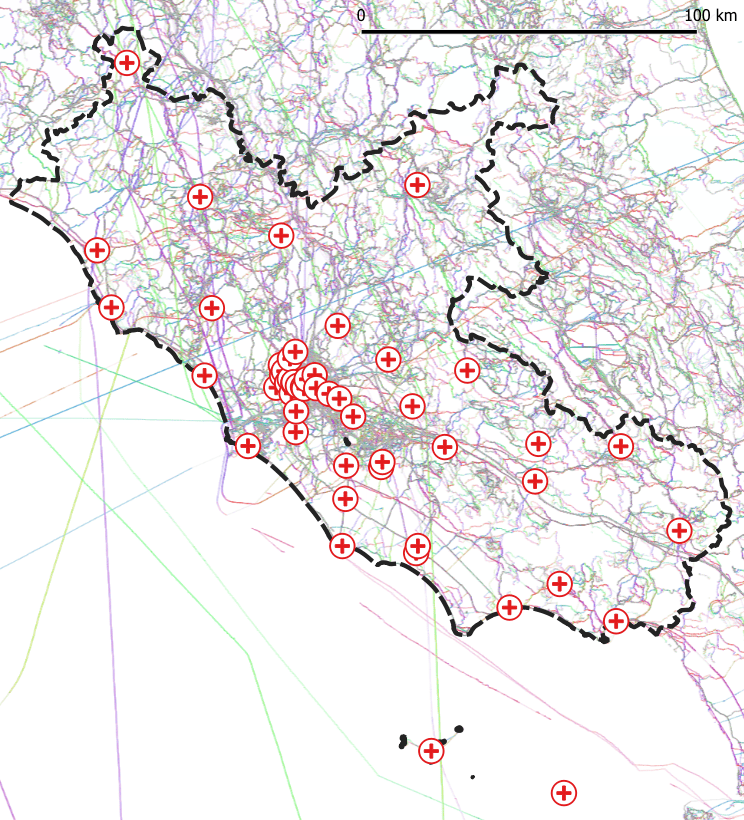}
	\caption{\bluemax{Map of the Lazio Region of Italy depicting all the EDs}.}
	\label{map allhospitals}
\end{figure}

	
	
	
\clearpage

\end{document}